\documentclass[traditabstract]{aa} 
\usepackage{txfonts}
\usepackage{graphicx}
\usepackage{longtable}
\usepackage{lscape}

\begin{document}

\title{The catalogue of radial velocity variable hot subluminous stars from the MUCHFUSS project}

\author{S.~Geier \inst{1}
   \and T.~Kupfer \inst{2}
   \and U.~Heber \inst{3}
   \and V.~Schaffenroth \inst{3,4}
   \and B.~N.~Barlow \inst{5}
   \and R.~H.~\O stensen \inst{6}
   \and S.~J.~O'Toole \inst{7}
   \and E.~Ziegerer \inst{3}
   \and C.~Heuser \inst{3}
   \and P.~F.~L.~Maxted \inst{8}
   \and B.~T.~G\"ansicke \inst{9}
   \and T.~R.~Marsh \inst{9}
   \and R.~Napiwotzki \inst{10}
   \and P.~Br\"unner \inst{3}
   \and M.~Schindewolf \inst{3}
   \and F.~Niederhofer \inst{1}}

\offprints{S.\,Geier,\\ \email{sgeier@eso.org}}

\institute{European Southern Observatory, Karl-Schwarzschild-Str.~2, 85748 Garching, Germany
\and Department of Astrophysics/IMAPP, Radboud University Nijmegen, P.O. Box 9010, 6500 GL Nijmegen, The Netherlands
\and Dr.~Karl~Remeis-Observatory \& ECAP, Astronomical Institute, Friedrich-Alexander University Erlangen-Nuremberg, Sternwartstr.~7, D 96049 Bamberg, Germany
\and Institute for Astro- and Particle Physics, University of Innsbruck, Technikerstr. 25/8, 6020 Innsbruck, Austria
\and Department of Physics, High Point University, 833 Montlieu Avenue, High Point, NC 27268, USA
\and Institute of Astronomy, KU Leuven, Celestijnenlaan 200D, B-3001 Heverlee, Belgium
\and Australian Astronomical Observatory, PO Box 915, North Ryde NSW 1670, Australia
\and Astrophysics Group, Keele University, Staffordshire, ST5 5BG, UK
\and Department of Physics, University of Warwick, Conventry CV4 7AL, UK
\and Centre of Astrophysics Research, University of Hertfordshire, College Lane, Hatfield AL10 9AB, UK}

\date{Received \ Accepted}

\abstract{
The project Massive Unseen Companions to Hot Faint Underluminous Stars from SDSS (MUCHFUSS) aims to find sdBs with compact companions like massive white dwarfs, neutron stars or black holes. Here we provide classifications, atmospheric parameters and a complete radial velocity (RV) catalogue containing 1914 single measurements for an sample of 177 hot subluminous stars discovered based on SDSS DR7. 110 stars show significant RV variability, while 67 qualify as candidates. We constrain the fraction of close massive compact companions {of hydrogen-rich hot subdwarfs} in our sample to be smaller than { $\sim1.3\%$}, which is already close to the theoretical predictions. However, the sample might still contain such binaries with longer periods exceeding $\sim8\,{\rm d}$. {We detect a mismatch between the $\Delta RV_{\rm max}$-distribution of the sdB and the more evolved sdOB and sdO stars, which challenges our understanding of their evolutionary connection.} Furthermore, irregular RV variations of unknown origin with amplitudes of up to $\sim180\,{\rm km\,s^{-1}}$ on timescales of years, days and even hours have been detected in some He-sdO stars. They might be connected to irregular photometric variations in some cases. 

\keywords{binaries: spectroscopic -- stars: subdwarfs -- stars: horizontal branch -- stars: atmospheres}}

\maketitle

\section{Introduction \label{sec:intro}}

Hot subdwarf stars (sdO/Bs) show spectral features similar to hot main sequence stars, but are much less luminous and therefore more compact. Depending on their spectral appearance, hot subdwarf stars can be divided into subclasses (Moehler et al. \cite{moehler90}; see Drilling et al. \cite{drilling13} for a more detailed classification scheme). While the observational classification seems straightforward, the formation and evolution of those objects is still unclear. 

In the Hertzsprung-Russell diagram most hot subdwarf stars are situated at the blueward extension of the Horizontal Branch (HB), the so called Extreme or Extended Horizontal Branch (EHB, Heber et al. \cite{heber86}). The most common class of hot subdwarfs, the sdB stars, are located on the EHB and are therefore considered to be core-helium burning stars. They have very thin hydrogen dominated atmospheres ($M_{\rm env}/M_{\rm sdB}\simeq10^{-3}$, $n_{\rm He}/n_{\rm H}\leq0.01$), their effective temperatures ($T_{\rm eff}$) range from $20\,000\,{\rm K}$ to $40\,000\,{\rm K}$ and their surface gravities ($\log{g}$) are one to two orders of magnitude higher than those of main sequence stars of the same spectral type (usually between $\log{g}=5.0$ and $6.0$). 

SdB stars are likely formed from stars that almost entirely lose their hydrogen envelopes after climbing up the red giant branch (RGB). The outer layer of hydrogen that remains does not have enough mass to sustain a hydrogen-burning shell, as is the case for cooler HB stars. Therefore the star can not evolve in the canonical way and ascend the Asymptotic Giant Branch (AGB). Instead  the star remains on the EHB until core-helium burning stops, and after a short time of shell-helium burning eventually reaches the white dwarf (WD) cooling tracks. According to evolutionary calculations the average lifetime on the EHB is of the order of $10^{8}\,{\rm yr}$ (e.g. Dorman et al. \cite{dorman93}). In this canonical scenario the hotter ($T_{\rm eff}=40\,000-80\,000\,{\rm K}$) and much less numerous hydrogen rich sdOs can be explained as rather short-lived shell-helium burning stars evolving away from the EHB.

Systematic surveys for radial velocity (RV) variations revealed that a large fraction of the sdB stars ($40-70\,\%$) are members of close binaries with orbital periods ranging from $\simeq0.05\,{\rm d}$ to $\simeq30\,{\rm d}$ (Maxted et al. \cite{maxted01}; Morales-Rueda et al. \cite{morales03}; Copperwheat et al. \cite{copperwheat11}). Most of the known companions of sdBs in radial velocity variable close binary systems are white dwarfs or late type main sequence stars, but substellar companions like brown dwarfs have been found as well (see Kupfer et al. \cite{kupfer15} and references therein). Those systems were most likely formed after a common envelope (CE) and spiral-in phase, which also provides an explanation for the required mass-loss on the RGB. However, apparently single sdBs and wide binary systems (Vos et al. \cite{vos12,vos13}; Barlow et al. \cite{barlow13}) exist as well. In those cases, it is less straightforward to explain the formation of the sdBs (see Geier \cite{geier13} for a review).

\begin{table}
\caption{\label{tab:instruments} Telescopes and instrumental setups}
\begin{center}
\begin{tabular}{llll}
\hline\hline
\noalign{\smallskip}
Telescope & Instrument & R & $\Delta \lambda$ [${\rm \AA}$] \\
\noalign{\smallskip}
\hline
\noalign{\smallskip}
Sloan & SDSS & $1800$ & $3800-9200$ \\
ESO-VLT & FORS1 & $1800$ & $3730-5200$ \\
WHT & ISIS & $4000$ & $3440-5270$ \\
CAHA-3.5m & TWIN & $4000$ & $3460-5630$ \\
ESO-NTT & EFOSC2 & $2200$ & $4450-5110$ \\
SOAR & Goodman & $2500$ & $3500-6160$\tablefootmark{a} \\
     & Goodman & $7700$ & $3700-4400$\\
Gemini & GMOS-N/S & $1200$ & $3770-4240$ \\
INT & IDS & $1400$ & $3000-6800$ \\
    & IDS & $4000$ & $3930-5100$\tablefootmark{b} \\
SAAO-1.9m & Grating & $4600$ & $4170-5030$ \\
\noalign{\smallskip}
\hline
\end{tabular}
\end{center}
\tablefoot{
\tablefoottext{a}{Used until 2011.}
\tablefoottext{b}{Additional data taken in March 2003 and April 2004.}
}
\end{table}

Hot subdwarf binaries with massive WD companions are good candidates for SN\,Ia progenitors. Due to gravitational wave radiation the orbit will shrink further and mass transfer from the sdB onto the WD will start once the sdB fills its Roche lobe. The Chandrasekhar limit might be reached either through He accretion on the WD (e.g. Yoon \& Langer \cite{yoon04} and references therein) or a subsequent merger of the system (Tutukov et al. \cite{tutukov81}; Webbink \cite{webbink84}). Two sdBs with massive WD companions have been identified to be good candidates for being SN\,Ia progenitors (Maxted et al. \cite{maxted00}; Geier et al. \cite{geier07}; Vennes et al. \cite{vennes12}; Geier et al. \cite{geier13b}). Neutron star (NS) or even black hole (BH) companions are predicted by theory as well (Podsiadlowski et al. \cite{podsi02}; Pfahl et al. \cite{pfahl03}). In this scenario two phases of unstable mass transfer are needed and the NS or the BH is formed in a supernova explosion. Nelemans (\cite{nelemans10}) showed that about 1\% of the short period sdBs should have NS companions whereas about 0.1\% should have BH companions. In an independent study Yungelson et al. (\cite{yungelson05}) predicted the number of systems with NS companions to be about 0.8\%. However, no NS/BH companion to an sdB has yet been detected unambiguously whereas a few candidates have been identified (Geier et al. \cite{geier10b}). Most recently, Kaplan et al. (\cite{kaplan13}) discovered the close companion to the pulsar PSR\,J1816+4510 to be a He-WD progenitor with atmospheric parameters close to an sdB star ($T_{\rm eff}=16\,000\,{\rm K}$, $\log{g}=4.9$).

The formation of the helium-rich classes of He-sdO/Bs is even more puzzling. Most (but not all) He-sdOs are concentrated in a very small region in the $T_{\rm eff}$-$\log{g}$ plane, slightly blueward of the EHB at $T_{\rm eff}=40\,000-80\,000\,{\rm K}$ and $\log{g}=5.60-6.10$ (Str\"oer et al. \cite{stroeer07}; Nemeth et al. \cite{nemeth12}). The He-sdBs are scattered above the EHB. The late hot flasher scenario provides a possible channel to form these objects (Lanz et al. \cite{lanz04}; Miller Bertolami et al. \cite{miller08}). After ejecting most of its envelope at the tip of the RGB, the stellar remnant evolves directly towards the WD cooling tracks and experiences a late core helium flash there. Helium and other elements like carbon or nitrogen are mixed into the atmosphere and the star ends up close to the helium main sequence. Depending on the depth of the mixing, stars with more or less helium in the atmospheres and different atmospheric parameters can be formed in this way. Most recently, Latour et al. (\cite{latour14}) found a correlation between the carbon and helium abundances of the He-sdOB stars in the globular cluster $\omega$~Cen, which is predicted by late hot flasher models. Hirsch (\cite{hirsch09a}) discovered a similar correlation for field helium-rich hot subdwarf (see also Heber \& Hirsch \cite{heber10}). Similar to the formation scenarios for sdB stars, the late hot flasher channel requires extreme mass-loss on the RGB probably triggered by binary interactions. However, the population of He-sdOs observed so far seems to consist mostly of single stars. Only one RV-variable He-sdO has been reported in the SPY sample, which corresponds to a fraction of only $3\,\%$ (Napiwotzki \cite{napiwotzki08}). However, higher fractions have been reported for the He-sdO populations in the PG sample (Green et al. \cite{green08}).

An alternative way of forming single hot subdwarfs is the merger of two helium white dwarfs in a close binary (Webbink \cite{webbink84}; Iben \& Tutukov \cite{iben84}). Loss of angular momentum through the emission of gravitational radiation will cause the system to shrink. Given the initial separation is small enough, the two white dwarfs eventually merge and if the mass of the merger is high enough, core-helium burning is ignited and a hot subdwarf is formed. Due to the strong mixing during the merger process, the atmospheres of the merger products are expected to be helium-rich (Zhang \& Jeffery \cite{zhang12}). 

Some hot subluminous stars are not connected to EHB-evolution at all. Objects with spectra and atmospheric parameters similar to normal sdBs are known, which are situated below the EHB (e.g. Heber et al. \cite{heber03}; Silvotti et al. \cite{silvotti12}). These objects are considered to be direct progenitors of helium white dwarfs, which descend from the red giant branch. For these {low-mass} post-RGB objects, which cross the EHB, evolutionary tracks indicate masses of about $0.20-0.33\,M_{\rm \odot}$ (Driebe et al. \cite{driebe98}). In order to form such objects, the mass loss on the RGB has to be more extreme than in the case of EHB stars. Objects down to even lower masses are known as extremely low-mass (ELM) WDs, which are members of close binary systems (e.g. Brown et al. \cite{brown12}). More massive He-stars, like the so-called low-gravity or luminous He-sdOs (Jeffery et al. \cite{jeffery08}) also belong to the class of hot subdwarfs and are situated between the EHB and the main sequence.

\section{The MUCHFUSS project} 

The project Massive Unseen Companions to Hot Faint Underluminous Stars from SDSS (MUCHFUSS) aims to find hot subdwarf stars with massive compact companions like massive white dwarfs ($>1.0\,M_{\rm \odot}$), neutron stars or stellar mass black holes. Hot subdwarf stars were selected from the Sloan Digital Sky Survey by colour and visual inspection of the spectra. Hot subdwarf stars with high radial velocity variations were selected as candidates for follow-up spectroscopy to derive the radial velocity curves and the binary mass functions of the systems. 

Geier et al. (\cite{geier11a}) discussed the target selection and the follow-up strategy. Detailed analyses of sdB binaries discovered in the course of this project are given in Geier et al. (\cite{geier11b}) and Kupfer et al. (\cite{kupfer15}). Three eclipsing systems have been discovered, two of them being the first sdBs with brown dwarf companions (Geier et al. \cite{geier11c}; Schaffenroth et al. \cite{schaffenroth14}). One system turned out to be the first sdB hybrid pulsator showing a reflection effect (\O stensen et al. \cite{oestensen13}). The photometric follow-up campaign of the MUCHFUSS project will be described in detail in Schaffenroth et al. (in prep). During dedicated spectroscopic MUCHFUSS follow-up runs bright sdB binary candidates were observed in a supplementary programme (Geier et al. \cite{geier13b,geier14}). Hot subdwarfs with a high but constant radial velocity were studied in the Hyper-MUCHFUSS project (Tillich et al. \cite{tillich11}).

Here we present classifications, radial velocities and atmospheric parameters of the close binary candidates discovered in the MUCHFUSS project so far. In Sect.~3 we describe the observations, target selection, classification and quantitative spectral analysis of our sample as well as the radial velocity catalogue. In Sect.~4 the different populations of RV variable hot subluminous stars are presented and discussed. A summary is then given in Sect.~5.

\section{Target selection, observations, spectroscopic analysis}

\begin{figure}[t!]
\begin{center}
	\resizebox{8.5cm}{!}{\includegraphics{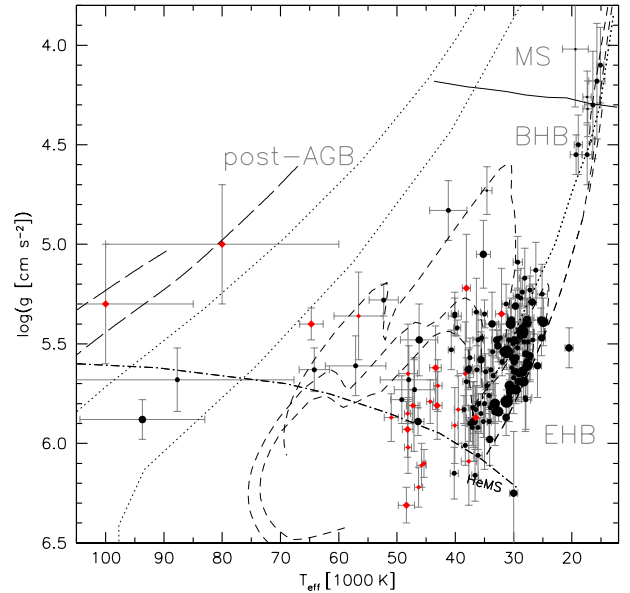}}
\end{center} 
\caption{$T_{\rm eff}-\log{g}$ diagram of the full sample of hot, subluminous, RV-variable stars. The size of the symbols scales with $\Delta RV_{\rm max}$. The black circles mark stars with hydrogen dominated atmospheres ($\log{y}<0$), while the red diamonds mark stars with helium dominated atmospheres. The helium main sequence (HeMS) and the HB band are superimposed with HB evolutionary tracks (dashed lines) for subsolar metallicity ($\log{z}=-1.48$) from Dorman et al. (\cite{dorman93}). The three tracks in the high temperature range correspond to helium core masses of $0.488$, $0.490$ and $0.495\,M_{\rm \odot}$ (from bottom-left to top-right). Those tracks mark the EHB evolution, since the stars do not reascend the giant branch in the helium shell-burning phase. The two tracks in the upper right correspond to core masses of $0.53$ and $0.54\,M_{\rm \odot}$. BHB stars following those tracks are expected to experience a second giant phase. The solid line marks the relevant part of the zero-age main sequence for solar metallicity taken from Schaller et al. (\cite{schaller92}). The two dotted lines are post-AGB tracks for hydrogen-rich stars with masses of $0.546$ (lower line) and $0.565\,M_{\rm \odot}$ (upper line) taken from Sch\"onberner (\cite{schoenberner83}). The two long-dashed lines are post-AGB tracks for helium-rich stars with masses of $0.53$ (lower line) and $0.609\,M_{\rm \odot}$ (upper line) taken from Althaus et al. (\cite{althaus09}).}
\label{tefflogg_full}
\end{figure}

\subsection{Observations and sample selection}

While the target selection presented in Geier et al. (\cite{geier11a}) includes SDSS up to Data Release 6 only, we have now applied the same selection criteria to Data Release 7 (Abazajian et al. \cite{abazajian09}). Hot subdwarf candidates were selected by applying a colour cut to SDSS photometry. All point source spectra within the colours $u-g<0.4$ and $g-r<0.1$ were selected and downloaded from the SDSS Data Archive Server\footnote{das.sdss.org}. By visual inspection we selected and classified $\simeq10\,000$ hot stars. Most objects much fainter than $g=19\,{\rm mag}$ have been excluded because of insufficient quality. The sample contains $1369$ hot subdwarfs, consistent with the preliminary number of hot subdwarfs ($1409$) found by Kleinman et al. (\cite{kleinman10}) in SDSS-DR7. 

The SDSS spectra are co-added from at least three individual integrations with typical exposure times of $15\,{\rm min}$ taken consecutively. We have obtained those individual spectra for stars brighter than $g=18.5\,{\rm mag}$. In addition, second epoch medium resolution spectroscopy was obtained from SDSS as well as our own observations, using ESO-VLT/FORS1, WHT/ISIS, CAHA-3.5m/TWIN and ESO-NTT/EFOSC2 (see Table~\ref{tab:instruments}, Geier et al. \cite{geier11a}). Typical exposure times ranged from $10\,{\rm min}$ to $20\,{\rm min}$. The S/N of the individual spectra ranges from about $15$ to about $100$.

{The radial velocities were measured by fitting a set of mathematical functions (Gaussians, Lorentzians and polynomials) to the spectral lines using the FITSB2 routine (Napiwotzki et al. \cite{napiwotzki04}). Three functions are used to match the continuum, the line and the line core, respectively and mimic the typical Voigt profile of spectral lines. 
The profiles are fitted to all suitable lines simultaneously using $\chi^{2}$-minimization and the RV shift with respect to the rest wavelengths with the associated $1\sigma$ error is measured. For the hydrogen-rich stars the Balmer and helium lines of sufficient strength have been used. For the helium-rich stars we used appropriate lines of neutral and single ionized helium. Since some of those stars still have significant hydrogen contamination we avoided the helium lines from the Pickering series, because they can be blended by the weaker hydrogen Balmer lines. Each single fit has been inspected visually and outliers caused by cosmic rays and other artifacts have been excluded. { Heliocentric corrections have been applied to the RVs and mid-JDs derived for the follow-up spectra, while the SDSS spectra available in the archive are already corrected.}

The average $1\sigma$ RV error of all the measurements in the catalogue is $\sim15\,{\rm km\,s^{-1}}$, which is consistent with independent checks of the SDSS wavelength stability using SDSS observations of F-stars ($<14.5\,{\rm km\,s^{-1}}$, Rebassa-Mansergas et al. \cite{rebassa07}). To correct for systematic shifts between different instruments we observed RV standards in our follow-up runs. The RMS scatter around the orbital fits of the solved binaries in our sample is also consistent with the formal uncertainties (for details, see Geier et al. \cite{geier11b}; Kupfer et al. \cite{kupfer15}).} We selected all objects with {maximum RV shifts} discrepant at the formal $1\sigma$-level and found $196$ candidates for RV variability. 

\begin{figure}[t!]
\begin{center}
	\resizebox{8.5cm}{!}{\includegraphics{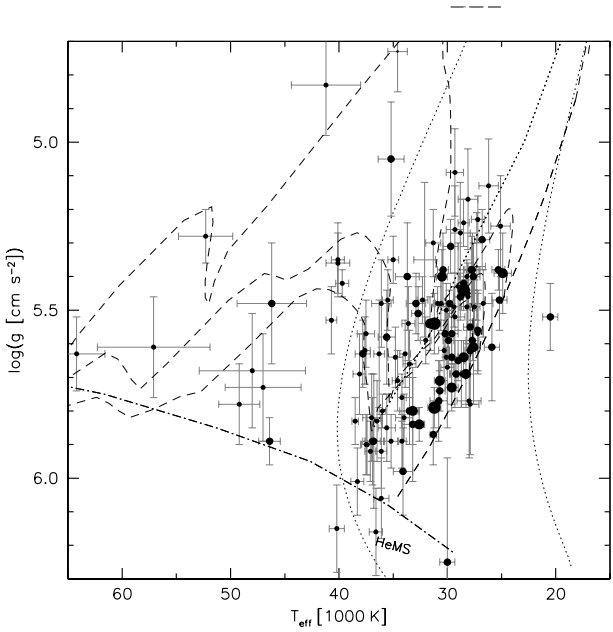}}
\end{center} 
\caption{$T_{\rm eff}-\log{g}$ diagram of RV variable hydrogen-rich sdB and sdOB stars (see Fig.~\ref{tefflogg_full}). The size of the symbols scales with $\Delta RV_{\rm max}$. The helium main sequence (HeMS) and the HB band are superimposed with HB evolutionary tracks (dashed lines) for subsolar metallicity ($\log{z}=-1.48$) from Dorman et al. (\cite{dorman93}). The three tracks correspond to helium core masses of $0.488$, $0.490$ and $0.495\,M_{\rm \odot}$ (from bottom-left to top-right). The two dotted lines mark post-RGB tracks (Driebe et al. \cite{driebe98}) for core masses of $0.234$ (left) and $0,259\,M_{\rm \odot}$ (right).}
\label{tefflogg_sdB}
\end{figure}

\subsection{Visual classification}

The basic classification of the hot subdwarf sample was done by visual inspection based on existence, width, and depth of helium and hydrogen absorption lines as well as the flux distribution between $4000$ and $6000\,{\rm \AA}$. Hot subdwarf B stars show strong and broad Balmer and weak (or no) He\,{\sc i} lines. SdOB stars show strong and broad Balmer lines as well as weak lines from He\,{\sc i} and He\,{\sc ii}, while sdO stars only display weak He\,{\sc ii} lines besides the Balmer lines. He-sdBs are dominated by strong He\,{\sc i} and sometimes weaker He\,{\sc ii} lines. Hydrogen absorption lines are shallow or not present at all. He-sdOs show strong He\,{\sc ii} and sometimes weak He\,{\sc i} lines. Balmer lines are not present or heavily blended by the strong He\,{\sc ii} lines of the Pickering series. A flux excess in the red compared to a reference spectrum as well as the presence of spectral features such as the Mg\,{\sc i} triplet at $5170\,{\rm \AA}$ or the Ca\,{\sc ii} triplet at $8650\,{\rm \AA}$ were taken as indications of a late-type companion.

From the total number of $1369$ hot subdwarfs, $983$ belong to the class of single-lined sdBs and sdOBs. Features indicative of a cool companion were found for $98$ of the sdBs and sdOBs. Nine sdOs show spectral features of cool companions, while $262$ sdOs, most of which show helium enrichment, are single-lined. 

{Comparing the results from the visual classification with the more detailed quantitative spectral analysis for the RV variable subsample presented here (see Sect.~\ref{sec:pop}), we conclude that our visual classification should be accurate to about $90\%$.} A catalogue with classifications and atmospheric parameters of the full SDSS sample including more recent data releases is in preparation. Here we restrict ourselves to the RV-variable sample. 

\subsection{Atmospheric parameters and spectroscopic distances}

To refine the visual classification and derive the atmospheric parameters a quantitative spectral analysis of the coadded SDSS spectra (or follow-up spectra of higher quality, if available) was performed for all RV variable stars in our sample with data of sufficient quality. The method is described in Geier et al. (\cite{geier11b}). We used appropriate model grids for the different classes of hot stars. The hydrogen-rich and helium-poor ($\log{y}=\log{n(\rm He)/n(\rm H)}<-1.0$) stars with effective temperatures below $30\,000\,{\rm K}$ were fitted using a of grid of metal line blanketed LTE atmospheres with solar metallicity. Helium-poor stars with temperatures ranging from $30\,000\,{\rm K}$ to $40\,000\,{\rm K}$ were analysed using LTE models with enhanced metal line blanketing (O'Toole \& Heber \cite{otoole06}). Metal-free NLTE models (Str\"oer et al. \cite{stroeer07}) were used for hydrogen-rich stars with temperatures below $40\,000\,{\rm K}$ showing moderate He-enrichment (log\,$y$\,=\,--1.0\,--\,0.0) and for hydrogen-rich sdOs. Finally, the He-sdOs were analysed with NLTE models taking into account the line-blanketing caused by nitrogen and carbon (Hirsch \& Heber \cite{hirsch09}). 

Spectroscopic distances to our stars have been calculated as described in Ramspeck et al. (\cite{ramspeck01}) assuming the canonical mass of $0.47\,{\rm M_{\odot}}$ for the subdwarfs {and appropriate masses for objects of other classes ($0.5\,M_{\rm \odot}$ for blue horizontal branch star candidates and $3.5\,M_{\rm \odot}$ for runaway main-sequence B star candidates,  Geier et al. \cite{geier15}; $0.6\,M_{\rm \odot}$ for post-AGB stars, Reindl et al. \cite{reindl15})} using the formula given by Lupton\footnote{http://www.sdss.org/dr6/algorithms/sdssUBVRITransform.html} to convert SDSS-g and r magnitudes to Johnson V magnitudes. Interstellar reddening was neglected in these calculations. 

\begin{figure*}[t!]
\begin{center}
	\resizebox{8.5cm}{!}{\includegraphics{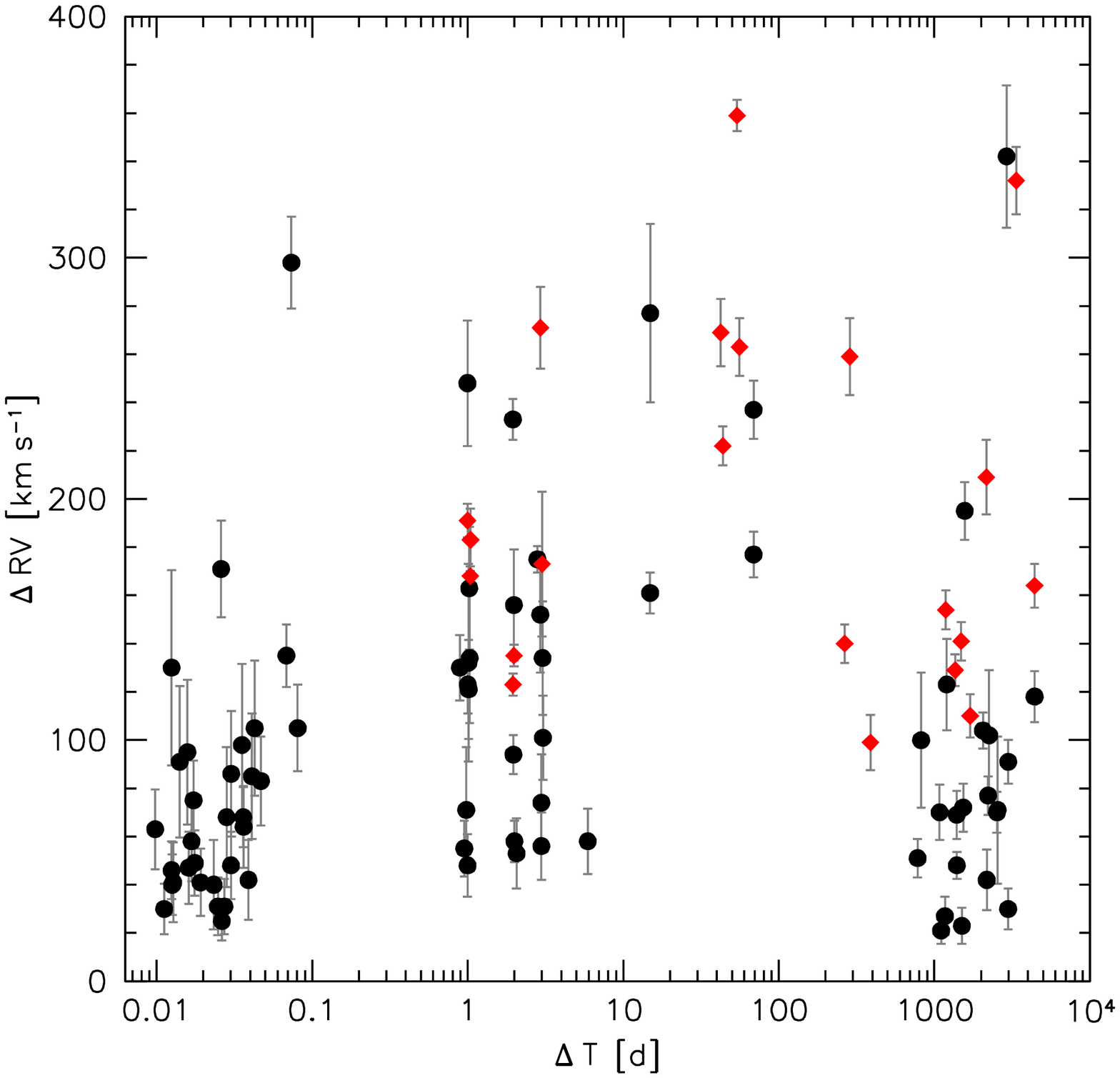}}
	\resizebox{8.5cm}{!}{\includegraphics{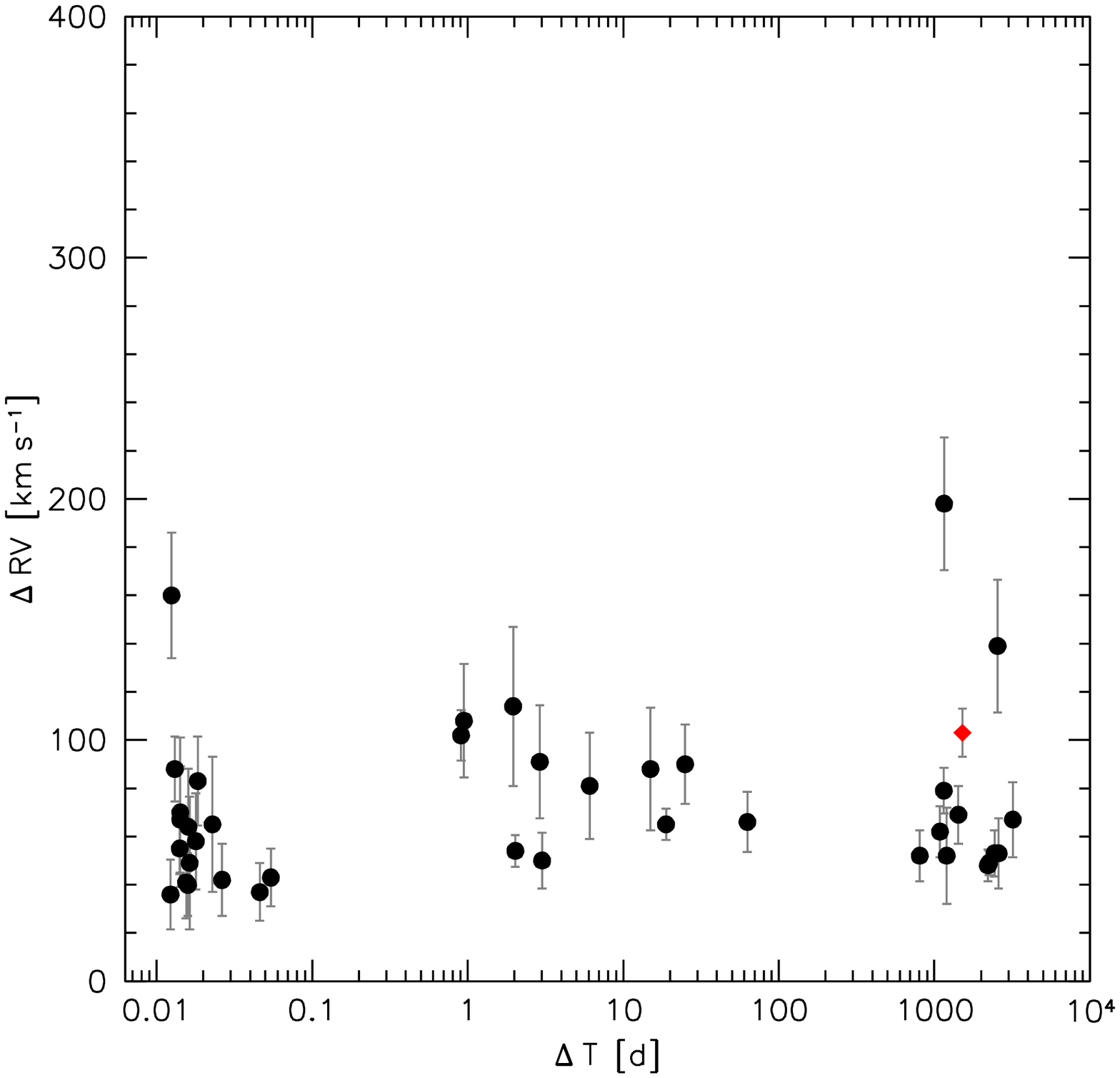}}
\end{center}
\caption{{\it Left panel:} Highest radial velocity shift between individual spectra plotted against time difference between the corresponding observing epochs. The filled red diamonds mark sdB binaries with known orbital parameters (Kupfer et al. \cite{kupfer15}), while the filled black circles mark the rest of the hydrogen-rich sdB sample of RV variable stars. {\it Right panel:} The same plot for the hydrogen-rich sdOB and sdO sample of RV variable stars.}
\label{dTdRV_ehb}
\end{figure*}

\subsection{Spectroscopic follow-up, criterion for variability and radial velocity catalogue}

During our follow-up campaign we obtained medium resolution ($R=1200-7700$), time-resolved spectroscopy { using WHT/ISIS, CAHA3.5m/TWIN, ESO-NTT/EFOSC2, SOAR/Goodman, Gemini/GMOS, INT/IDS and the grating spectrograph at the 1.9m telescope at SAAO} (see Table~\ref{tab:instruments}, Geier et al. \cite{geier11b}; Kupfer et al. \cite{kupfer15}) and measured the RVs as described above. 

To estimate the fraction of false detections produced by random fluctuations and calculate the significance of the measured RV variations we apply the method outlined in Maxted et al. (\cite{maxted01}). For each star we calculate the inverse-variance weighted mean velocity from all measured epochs. Assuming this mean velocity to be constant, we calculate the $\chi^{2}$. Comparing this value with the $\chi^{2}$-distribution for the appropriate number of degrees of freedom we calculate the probability $p$ of obtaining the observed value of $\chi^{2}$ or higher from random fluctuations around a constant value. The maximum RV shifts ($\Delta RV_{\rm max}$), the average $1\sigma$ uncertainties of the two corresponding measurements, the timespan between those two epochs and the logarithm of the false-detection probability $\log{p}$ are given in Tables~\ref{tab1}-\ref{tab3}. 

We consider the detection of RV variability to be significant, if the false-detection probability $p$ is smaller than $0.01\%$ ($\log{p}<-4.0$). The fraction of such significant detections in our initial sample of 196 is $56\%$ (110 objects). Objects with false-detection probabilities between $0.01\%$ and $5\%$ ($\log{p}=-4.0$ to $\log{p}=-1.3$) are regarded as candidates for RV variability and constitute $34\%$ of the initial sample (67 objects). About $10\%$ ($\log{p}>-1.3$, 19 objects) are regarded as non-detections (the parameters of those stars can be found in Table~\ref{app:tab1}). {Removing those non-detections we end up with a sample of 177 stars, which show RV variability with probabilites between $95\%$ and $99.9\%$ (see Table~\ref{tab:summary}).} Orbital solutions were already derived for 22 close binary sdB systems (see Kupfer et al. \cite{kupfer15} and references therein).

The catalogue contains $1914$ epochs (mid-HJD), associated radial velocities and $1\sigma$-RV-uncertainties of the RV variable stars as well as information about the instruments used to obtain the spectra. It can be accessed online from the Vizier database operated by CDS. 

\begin{table}
\caption{\label{tab:summary} Sample statistics}
\begin{center}
\begin{tabular}{llll}
\hline\hline
\noalign{\smallskip}
Class & RV variable & RV variable & non- \\
      &             & candidates  & detections \\
\noalign{\smallskip}
\hline
\noalign{\smallskip}
H-rich sdO/B  &  89  &  50  &  13 \\
He-rich sdO/B &  14  &  11  &  4 \\
Others        &   7  &   6  &  2 \\
\noalign{\smallskip}
\hline
\noalign{\smallskip}
Total         & 110  & 67   &  19 \\
\noalign{\smallskip}
\hline
\end{tabular}
\end{center}
\end{table}

\section{Radial velocity variable populations of hot subluminous stars}\label{sec:pop}

Since our sample has been preselected in the way outlined above it is not straightforward to derive the true fractions of RV variable stars for each class of hot subdwarfs. Most stars in our sample have been selected based on RV variations between the individual SDSS spectra, which have usually been taken within just $45\,{\rm min}$. Only binaries with sufficiently short orbital periods and high RV amplitudes show significant variations on such short timescales, while binaries with smaller RV amplitudes and longer periods remain undetected. 

Fig.~\ref{tefflogg_full} shows the $T_{\rm eff}-\log{g}$-diagram of the RV variable sample. Most of the stars are indeed associated with the EHB and therefore most likely core or shell-helium burning hot subdwarfs. Four objects have higher temperatures and are more likely hydrogen and helium-rich post-AGB objects. Nine stars have temperatures below $20\,000\,{\rm K}$ and most of them are likely associated with the blue horizontal branch (see Table~\ref{tab3}). The B-type binary candidates are discussed separately in Geier et al. (\cite{geier15}), the hot post-AGB stars in Reindl et al. (\cite{reindl15}).

Although only the orbits of 22 binaries from our sample have been solved, the distribution of $\Delta RV_{\rm max}$ can be used as a diagnostic tool as well. The width of this distribution scales with the width of the companion mass distribution as well as the distribution of orbital periods. 

\subsection{Hydrogen-rich hot subdwarf stars and their evolutionary connection} 

The most common class of RV variable objects in our sample are sdB, sdOB and sdO stars with hydrogen-dominated atmospheres (see Table~\ref{tab1}). Fig.~\ref{tefflogg_sdB} shows the $T_{\rm eff}-\log{g}$-diagram of this subsample. As expected, most objects are concentrated on the EHB and some objects follow the tracks of more evolved shell-helium burning stars. This distribution is consistent with other studies (e.g. Nemeth et al. \cite{nemeth12}). However, it is not clear whether all objects situated above the EHB are really shell-helium burning stars that evolved along the predicted evolutionary tracks. Other objects like {low-mass} post-RGB stars evolve in a different way and might constitute a certain fraction of stars in this region of the $T_{\rm eff}-\log{g}$-diagram.

{ The detected RV-variability in those objects is very likely caused by binary motion. Up to now the orbital parameters of 142 close binaries have been measured. Most of the solved systems have hydrogen-rich sdB primaries, but this sample also contains 46 hydrogen-rich sdOB and sdO stars (see Kupfer et al. \cite{kupfer15} and references therein, but see also the discussion in Sect.~\ref{sec:hesdo}). Another possible source of RV-variations are short-period p-mode pulsations. However, the fraction of pulsating hot subdwarf stars is quite small (about $5\%$) and the RV-variations are usually smaller than our detection limit. Even in the most extreme cases known, where those variations can reach amplitudes of $10-20\,{\rm km\,s^{-1}}$ (e.g. O'Toole et al. \cite{otoole05}), we would most likely not resolve and detect them in our data, because our exposure times are usually longer than the typical periods (a few minutes) of those pulsations.}

The additional information provided by the RV variability (Table~\ref{tab1}, Fig.~\ref{dTdRV_ehb}) allows us to probe the connection between objects on the EHB classified as sdBs (100 RV variable objects) with stars that are situated above the EHB classified as sdOB or sdO (40 RV-variable objects). While the internal structure and the atmospheric parameters of the hot subdwarf change with time, the orbital period and the companion mass are not predicted to change significantly within the lifetime of the sdB ($\sim100\,{\rm Myr}$). A significant shortening of the orbital period due to angular momentum lost by gravitational wave emission is only predicted for the most compact binaries with the most massive companions, which turned out to be quite rare (e.g. Geier et al. \cite{geier07,geier13b}). Furthermore, the orbital evolution will always lead to shorter periods and therefore higher RV-amplitudes. If the sdBs on the EHB evolve to become hydrogen-rich sdOB and sdO stars, the $\Delta RV_{\rm max}$-distribution should esentially remain the same.
 
\begin{figure}[t!]
\begin{center}
	\resizebox{8.5cm}{!}{\includegraphics{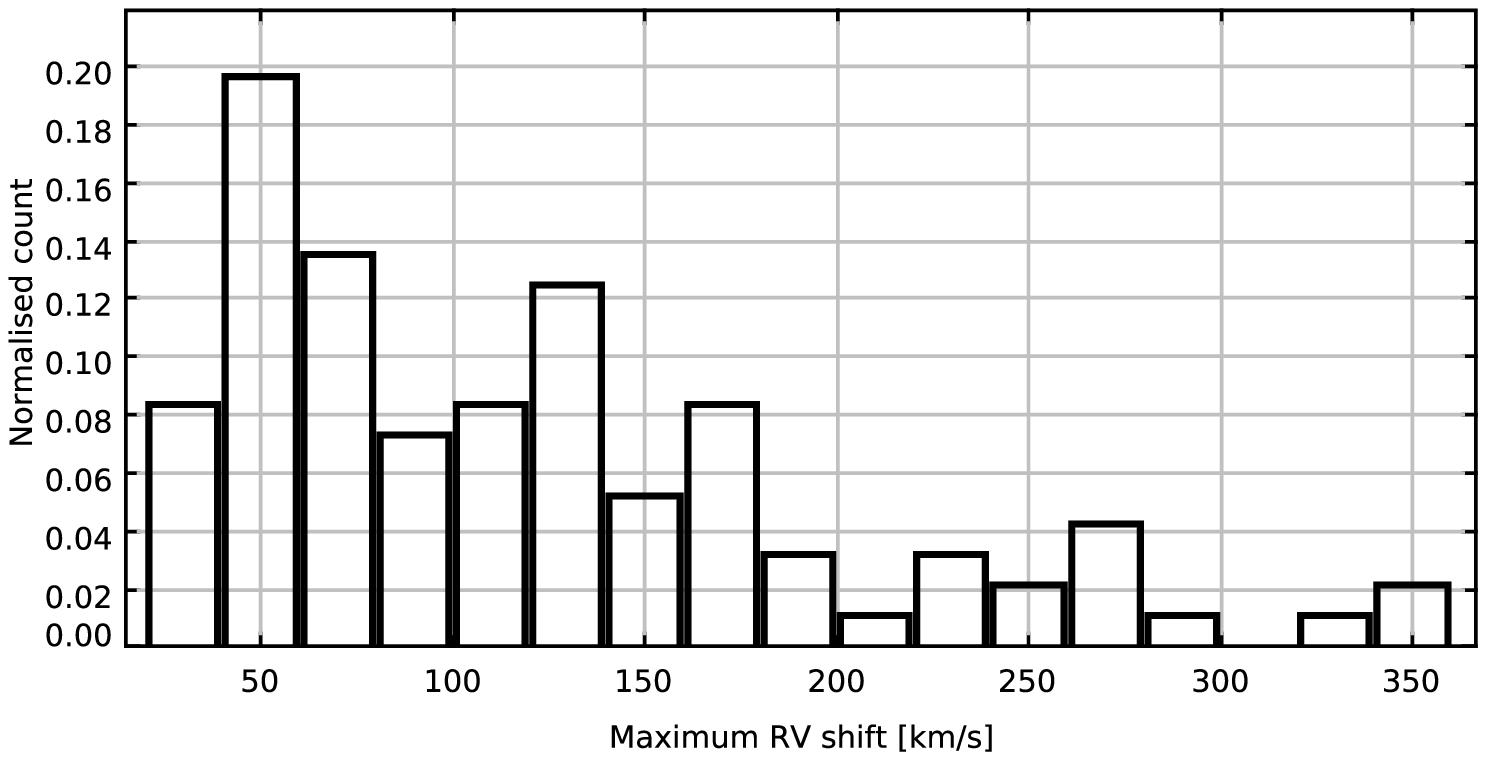}}
	\resizebox{8.5cm}{!}{\includegraphics{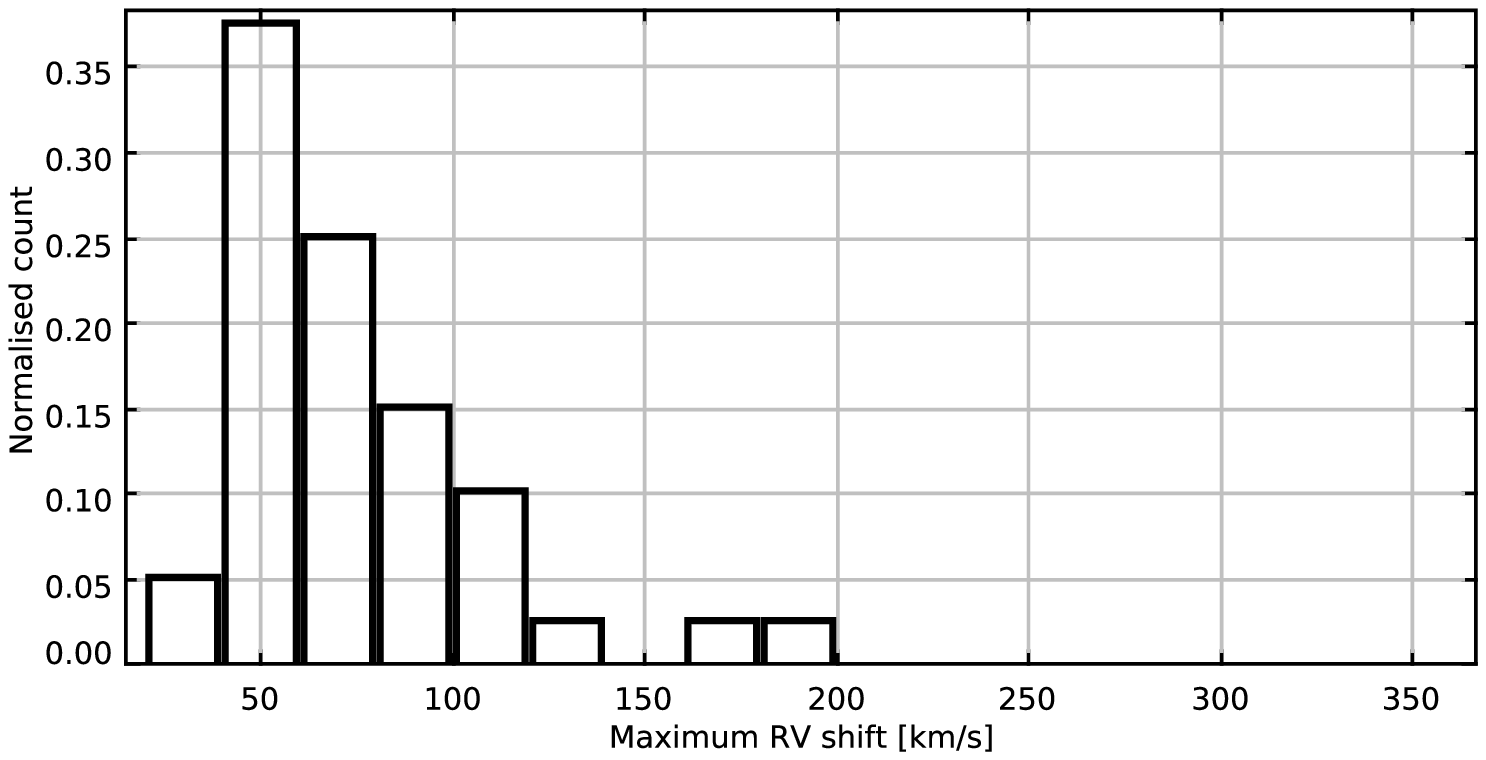}}
\end{center}
\caption{$\Delta RV_{\rm max}$ distribution of RV-variable sdB stars (upper panel) as well as sdOB and sdO stars with hydrogen-rich atmospheres (lower panel).}
\label{dRV_hist_sdb}
\end{figure}

\begin{figure}[t!]
\begin{center}
	\resizebox{8.5cm}{!}{\includegraphics{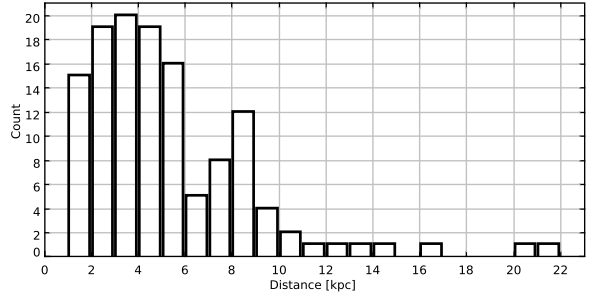}}
\end{center}
\caption{Distribution of spectroscopic distances for the hydrogen-rich sdB, sdOB and sdO stars (see Table~\ref{tab1}).}
\label{dist_sdb}
\end{figure}

Fig.~\ref{dRV_hist_sdb} shows the $\Delta RV_{\rm max}$-distribution of both subsamples. While the distribution below $100\,{\rm km\,s^{-1}}$ looks very similar as expected, the sdB sample  shows a wider range of RV shifts { (see also Fig.~\ref{dTdRV_ehb})}. This can only be partly explained by the smaller size of the sdOB/sdO sample. {There are also no significant differences in data quality and temporal sampling between the two different groups. Essentially the same hydrogen Balmer lines have been used to measure the RVs. The reason for this mismatch, which challenges our understanding of EHB evolution, is unclear.}

\subsection{Low-mass post-RGB binaries}

Our sample contains two sdBs that might be good candidates for {low-mass} post-RGB stars. With a low effective temperature of $20\,500\,{\rm K}$ and a rather high surface gravity $\log{g}=5.52$ J083334.76-045759.4 is situated well below the EHB. Such a location is inconsistent with core-helium burning. Furthermore, it shows a high $\Delta RV_{\rm max}=161\,{\rm km\,s^{-1}}$. J094750.71+162731.8 is hotter ($T_{\rm eff}=30000\,{\rm K}$), but has a very high surface gravity $\log{g}=6.25$. Also situated below the EHB it shows $\Delta RV_{\rm max}=130\,{\rm km\,s^{-1}}$. However, whether a significant contribution of {low-mass} post-RGB binaries leads to the wider distribution of RV-shifts, still needs to be studied in more detail (see also discussion in Geier et al. \cite{geier13a}).

\subsection{Hierarchical triple systems} 

One sdB in our sample is a double-lined system and shows weak spectral features of a main-sequence companion. J205101.72+011259.7 shows a shift of $91.0\pm31.5\,{\rm km\,s^{-1}}$ within just $0.0141\,{\rm d}$ with a false-detection probability of only $0.05\%$. It is very unlikely that this variation is caused by the main-sequence companion. The solved orbits of sdB+MS binaries have long periods of the order of $1000\,{\rm d}$ (Vos et al. \cite{vos12,vos13}; Barlow et al. \cite{barlow12,barlow13}). We therefore conclude that J205101.72+011259.7 is another candidate for a hierarchical triple system consisting of an sdB in a short-period binary with unseen companion and a main sequence star orbiting this inner binary with a long period (e.g. Barlow et al. \cite{barlow14}, see also discussion in Kupfer et al. \cite{kupfer15}). 

\subsection{The fraction of massive compact companions}

The primary aim of the MUCHFUSS project is to find massive compact companions to sdB stars. However, only two known sdB binaries with periods shorter than $0.1\,{\rm d}$ and $\Delta RV_{\rm max}\sim700\,{\rm km\,s^{-1}}$ have WD companions with masses exceeding $0.7\,M_{\rm \odot}$ (Geier et al. \cite{geier07,geier13b}). SdB+NS/BH binaries with similar periods would have $\Delta RV_{\rm max}>1000\,{\rm km\,s^{-1}}$. 

However, the highest $\Delta RV_{\rm max}$ measured in our subsample of hydrogen-rich sdB, sdOB and sdO stars is just $359\,{\rm km\,s^{-1}}$ (see Table~\ref{tab1}). Due to the RV sampling of our dataset provided by the individual SDSS spectra it is very unlikely that we have missed a short-period {($0.1\,{\rm d}$)} binary with an $\Delta RV_{\rm max}>1000\,{\rm km\,s^{-1}}$ by chance. To estimate an upper limit for the fraction of {such extremely close} binary sdB+NS/BH binaries in our sample we count the number of { hydrogen-rich sdBs and sdOBs with significant RV variability ($\log{p}<-4.0$) in our sample (76 objects, see Table~\ref{tab1})} and invert it. In this way we derive the fraction of those objects in our sample to be smaller than { $1.3\%$}. This fraction is still consistent with the theoretically predicted fractions of about $1\%$ (Yungelson et al. \cite{yungelson05}; Geier et al. \cite{geier10b}; Nelemans \cite{nelemans10}).

{However, we would most likely not expect the most massive compact companions in our sample to have such short orbital periods anyway.} To allow the massive companion to spiral in deep enough to form such compact binaries during the common envelope phase, the red-giant progenitors of the sdB stars are predicted to have tightly bound envelopes and to be rather massive ($2-3\,M_{\rm \odot}$, Geier et al. \cite{geier13b}). Such stars are only found in young field populations like the Galactic thin disk and the two sdB binaries with the most massive WD companions known so far indeed belong to this population (Maxted et al. \cite{maxted00}; Geier et al. \cite{geier07}; Geier et al. \cite{geier13b}). 

\begin{figure}[t!]
\begin{center}
	\resizebox{8.5cm}{!}{\includegraphics{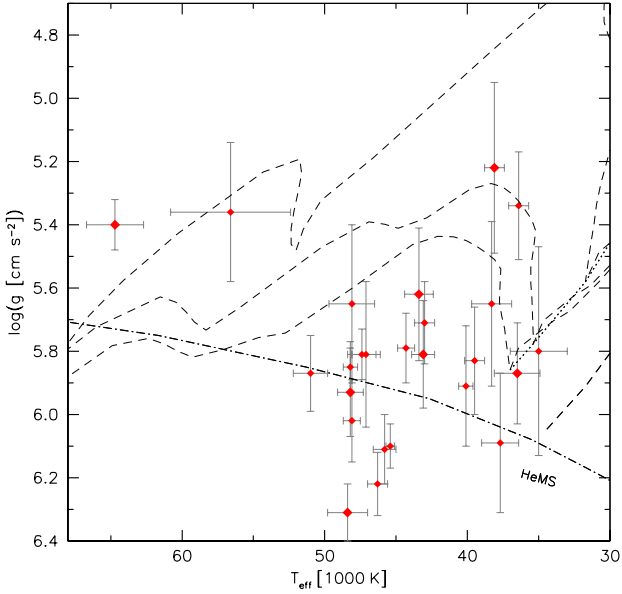}}
\end{center} 
\caption{$T_{\rm eff}-\log{g}$ diagram of RV variable helium-rich sdOB and sdO stars (see Fig.~\ref{tefflogg_full}). The size of the symbols scales with $\Delta RV_{\rm max}$. The helium main sequence (HeMS) and the HB band are superimposed with HB evolutionary tracks (dashed lines) for subsolar metallicity ($\log{z}=-1.48$) from Dorman et al. (\cite{dorman93}). The three tracks  correspond to helium core masses of $0.488$, $0.490$ and $0.495\,M_{\rm \odot}$ (from bottom-left to top-right).}
\label{tefflogg_hesdO}
\end{figure} 

Fig.~\ref{dist_sdb} shows the distribution of spectroscopic distances for the sample. Those distances range from $1$ to more than $20\,{\rm kpc}$. Taking into account that the SDSS footprint mostly covers high Galactic latitudes and assuming a scale-height of $\sim0.3\,{\rm kpc}$ for the thin disk, we conclude that the vast majority of the stars in our sample do not belong to this young population. Most binary candidates exceeding $d\sim3\,{\rm kpc}$ should belong to the old Galactic halo population, the rest to the intermediate thick disk population. Since both populations do not contain intermediate mass main-sequence stars, which are the likely progenitors of short-period sdBs with massive compact companions, it is no surprise that we do not find them in our sample.

\begin{figure}[t!]
\begin{center}
	\resizebox{8.5cm}{!}{\includegraphics{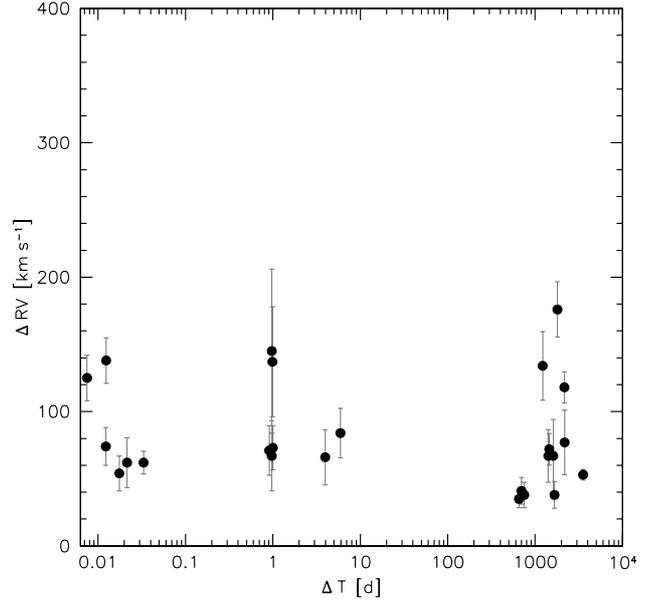}}
\end{center}
\caption{Highest radial velocity shift between individual spectra plotted against time difference between the corresponding observing epochs for helium-rich sdO and sdOB stars (see Fig.~\ref{dTdRV_ehb}).}
\label{dTdRV_hesdo}
\end{figure} 

While we can exclude sdB binaries {with periods of a few hours and massive compact companions}, our sample might still contain such objects with longer periods. Since more massive companions are expected  to be quite efficient in ejecting the common envelope, such binaries might exist. Taking into account the $\Delta RV_{\rm max}$-distribution and the fraction of  {solved binary orbits} (see Fig.~\ref{dTdRV_ehb}), we estimate that a {yet undetected} population of long-period binaries with $K<100\,{\rm km\,s^{-1}}$ might be present. Assuming the canonical sdB mass of $0.47\,M_{\rm \odot}$ and a minimum companion mass at the Chandrasekhar limit ($1.4\,M_{\rm \odot}$) this translates into orbital periods longer than $\sim8\,{\rm d}$.

\begin{figure*}[t!]
\begin{center}
	\resizebox{8.5cm}{!}{\includegraphics{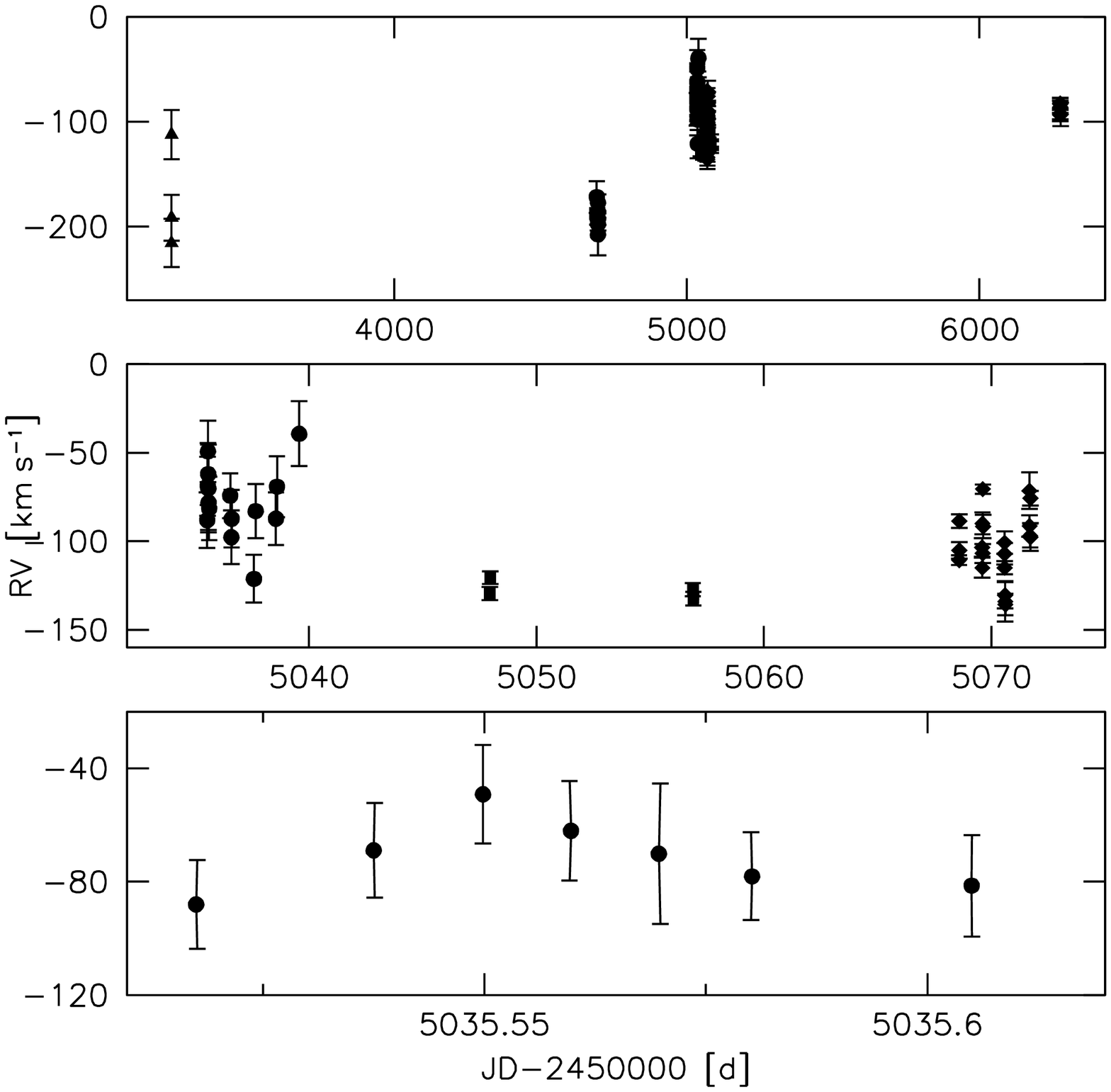}}
	\resizebox{8.5cm}{!}{\includegraphics{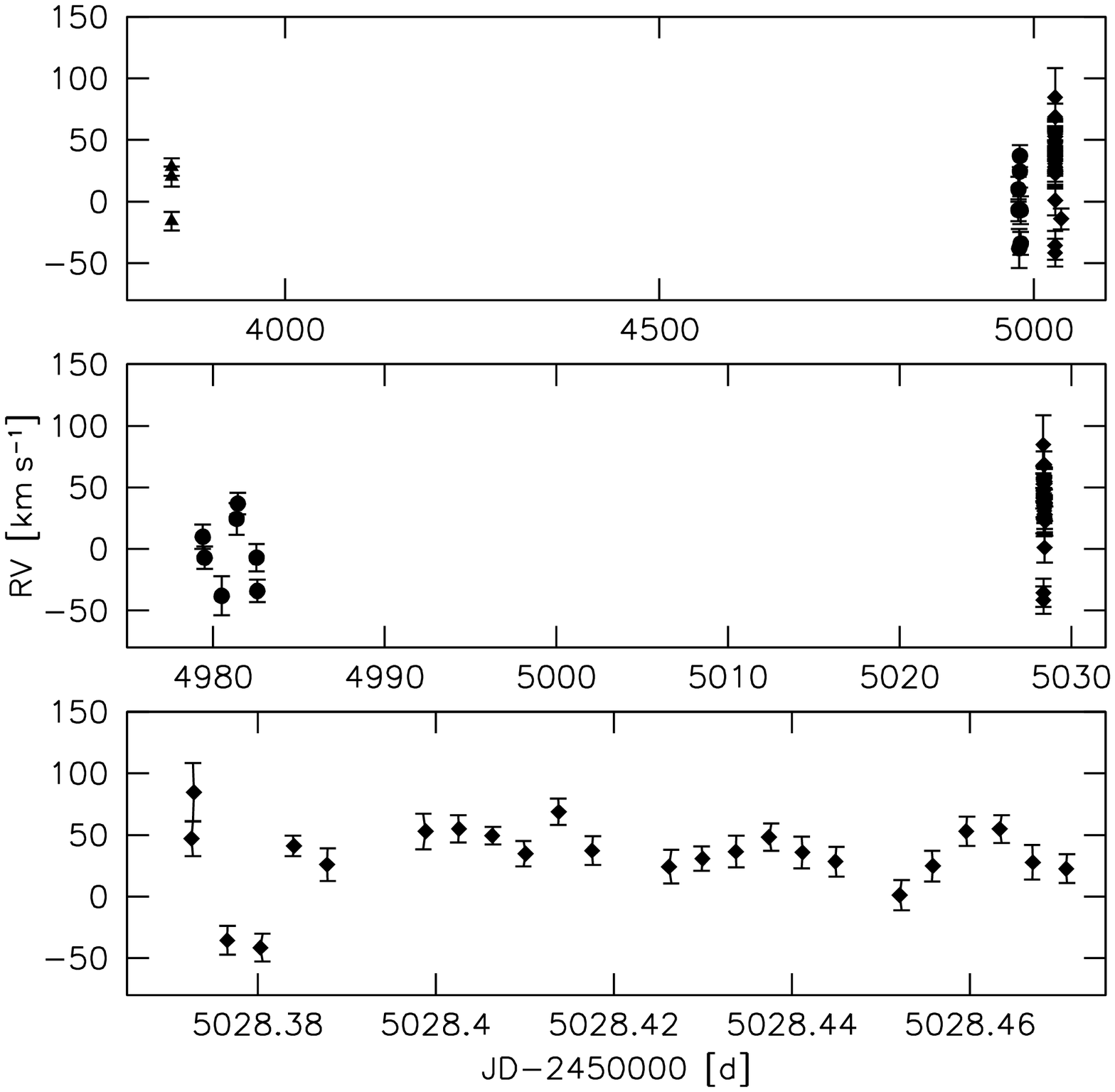}}
\end{center}
\caption{Radial velocities of J232757.46+483755.2 (left panels) and J141549.05+111213.9 (right panels) against Julian date. Significant variations are present on timescales of years (upper panels), days (middle panels) and hours (lower panels).}
\label{RV_var}
\end{figure*}

\subsection{Irregular RV variations of helium-rich hot subdwarf stars}\label{sec:hesdo}

Our RV-variable sample contains $29$ helium-rich hot subdwarf stars. $14$ of them show significant RV variations while $15$ qualify as candidates (see Table~\ref{tab2}). Most of them are situated close to the He-MS in the $T_{\rm eff}-\log{g}$-diagram (see Fig.~\ref{tefflogg_hesdO}) and the atmospheric parameters are quite typical for the field population of He-sdOs ($T_{\rm eff}=40\,000-50\,000\,{\rm K}$, Str\"oer et al. \cite{stroeer07}; Nemeth et al. \cite{nemeth12}). However, quite a number of stars have lower temperatures between $35\,000\,{\rm K}$ and $40\,000\,{\rm K}$. Those helium-rich sdOBs are rare in the field population, but quite dominant in the globular cluster $\omega$~Cen (Latour et al. \cite{latour14}). Following the discussion in Latour et al. (\cite{latour14}) this might be related to the age of the parent population, since most of the stars in our sample belong to the old thick disk or halo populations, while most of the bright stars studied by Nemeth et al. (\cite{nemeth12}) belong to the young thin disk population. The He-sdOs J232757.46+483755.2 and J110215.45+024034.1 seem to be more evolved than the rest of the sample and might also be associated to the helium-rich post-AGB stars. 

J160450.44+051909.2 and J160623.21+363005.4 belong to the class of He-sdOBs with lower surface gravity (e.g. Naslim et al. \cite{naslim10}).\footnote{In the literature those objects are usually called He-sdBs, but here we follow the more detailed spectroscopic classification outlined in Sect.~3.2.} Only one He-sdOB is known to be in a close double-lined, spectroscopic binary with an almost identical companion of the same type (Sener \& Jeffery \cite{sener14}). Another close binary contains an sdB with intermediate helium-enrichment 
(Naslim et al. \cite{naslim12}). Follow-up observations are needed to study the binary properties of those rare objects and compare them with the other hot subdwarf 
populations.  

The discovery of RV variable He-sdOs (Green et al. \cite{green08}; Geier et al. \cite{geier11a}) on the other hand seemed to be inconsistent with the idea, that those stars 
are formed by He-WD mergers (e.g. Webbink \cite{webbink84}), because merger products are expected to be single stars. Fig.~\ref{dTdRV_hesdo} shows the maximum RV shifts 
between individual spectra plotted against the time difference between the corresponding epochs. When compared with Fig.~\ref{dTdRV_ehb} one can see that there are no stars with shifts higher than $\sim200\,{\rm km\,s^{-1}}$ and that the number of objects showing shifts at short timespans ($<0.1\,{\rm d}$) is smaller as well. 

Because of the important implications for their formation, we were eager to solve the first He-sdO binaries and gave them high priority in our follow-up campaign. However, although 
we gathered up to $59$ epochs for some of them, we were not able to find a single orbital solution. Adding more data in general degraded preliminary solutions that looked 
promising. Besides assuming circular orbits we also allowed for eccentricity and explored especially the parameter space of high orbital eccentricities (see Geier et al. 
\cite{geier11b}). No periodic variations could be detected with sufficient significance. 

Fig.~\ref{RV_var} shows the radial velocities of the two He-sdO stars J141549.05+111213.9 and J232757.46+483755.2 for which we obtained the most data. Significant RV variations with amplitudes of up to $\sim180\,{\rm km\,s^{-1}}$ are seen on timescales of years, days and even hours. While J141549.05+111213.9 has atmospheric parameters typical for He-sdO stars, J232757.46+483755.2 has a higher effective temperature and seems to be more evolved (see Table~\ref{tab1}, Fig.~\ref{tefflogg_hesdO}).

\begin{figure}[t!]
\begin{center}
	\resizebox{8.5cm}{!}{\includegraphics[angle=-90]{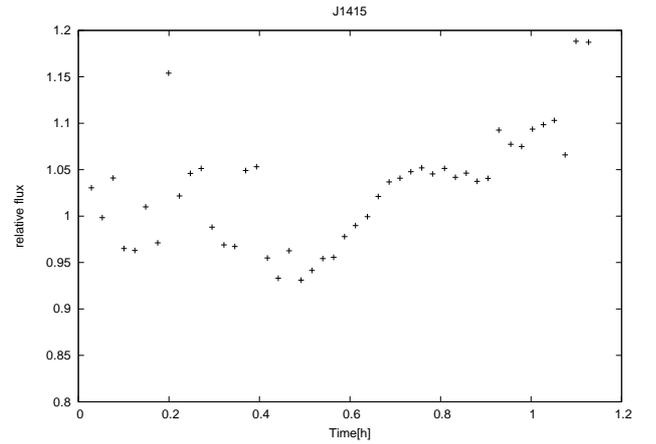}}
\end{center}
\caption{$V_{\rm B}$-band lightcurve of J141549.05+111213.9 taken with the BUSCA camera mounted at the 2.2m telescope at Calar Alto (Schaffenroth et al. in prep.).}
\label{J1415_lc}
\end{figure}

The origin of these irregular RV variations remains unclear. \O stensen et al. (\cite{oestensen10}) reported the discovery of irregular variations in the light curve of the He-sdOB 
star J19352+4555 observed by the Kepler mission. Jeffery et al. (\cite{jeffery13}) found another He-sdOB star with Kepler light curves (KIC\,10449976) that shows a variation with a period of $3.9\,{\rm d}$ and variable amplitude. Radial velocity follow-up with time-resolved spectroscopy revealed a possible, but still marginal RV variability of $50\pm20\,{\rm km\,s^{-1}}$. Most recently, Green et al. (\cite{green14}) reported irregular variations in the lightcurves of two helium-rich and one hydrogen-rich sdO star. They also found RV variations of up to $20\,{\rm km\,s^{-1}}$ for some hydrogen- and helium-rich sdO stars. 

During our photometric follow-up campaign (Schaffenroth et al. in prep.) we obtained a light curve of J141549.05+111213.9 (see Fig.~\ref{J1415_lc}) showing irregular variations very similar to the ones found by Green et al. (\cite{green14}). There might therefore be a link between those two phenonema. A similar light curve of J232757.46+483755.2 showed no such variations, but this might also be an indication for their transient nature. It remains to be seen whether the high RV variations we found are really connected to the light curve variations observed in similar stars. 

{ Whether this behaviour is restricted to helium-rich sdOs only or might also affect hydrogen-rich sdOs is unclear. The possibly connected photometric variations discovered by Green et al. (\cite{green14}) affect both kinds of sdOs. Also the distributions of maximum RV-shifts for both populations look quite similar (see Fig.~\ref{dTdRV_ehb} right panel, Fig.~\ref{dTdRV_hesdo}). However, since we focused our follow-up mostly on hydrogen-rich sdBs and helium-rich sdOs, we did not obtain a sufficient number of RVs to check for irregular variations in one of the hydrogen-rich sdOs. An important difference between the two populations is that at least some hydrogen-rich sdOs are known to reside in close binaries (see Kupfer et al. \cite{kupfer15} and references therein), whereas not a single He-sdO in a close binary system has been found yet.}

Some ideas have been put forward to explain the light curve variations. Jeffery et al. (\cite{jeffery13}) suggested that the variations might be due to star spots caused by magnetic fields. They also discuss the possibility of a shallow reflection effect originating from the irradiated hemisphere of a cool low-mass companion. Bear \& Soker (\cite{bear14}) propose a heated planetary companion with strong weather to be responsible for the variability of KIC\,10449976. Green et al. (\cite{green14}) see strong similarities of the variations detected in their stars to the variations seen in some cataclysmic variables and attribute them to the presence of accretion disks. The high and irregular RV variations seen in our objects can be hardly explained in those ways. A reflection effect binary with a low-mass companion should show periodic variations with small RV amplitudes and the presence of an accretion disk would require a close companion as well. 

{Another possible reason might be magnetic activity of those stars. Heber et al. (\cite{heber13}) reported the discovery of a He-sdO star with significant Zeeman-splitting and a magnetic field of several hundred kG. More of those objects have been discovered recently (Nemeth priv. comm.). Variable magnetic fields might lead to distortions of the spectral lines, which are not resolved in the medium-resolution spectra we have and may mimic irregular RV shifts. However, the very high RV shifts observed are again hardly consistent with such a scenario. High-resolution, time-resolved follow-up photometry, spectroscopy, and maybe also spectropolarimetry are necessary to study those mysterious RV shifts.}

\section{Summary}

In this paper we provide classifications, atmospheric parameters and a complete RV catalogue containing 1914 single measurements of the 177 {most likely} RV variable hot subluminous stars discovered in the MUCHFUSS project from SDSS DR7. 

We detect a mismatch between the $\Delta RV_{\rm max}$-distribution of the sdB and the more evolved sdOB and sdO stars, {which challenges our understanding of their evolutionary connection. Our sample contains two candidates for He-WD progenitors.} Furthermore, one of the RV variable sdB binaries is double-lined and a candidate for a hierarchical triple system.

Based on the $\Delta RV_{\rm max}$-distribution of the { hydrogen-rich sdB and sdOB stars we constrain the fraction of close massive compact companions in our sample to be smaller than $\sim1.3\%$.} However, the sample might still contain such binaries with longer periods exceeding $\sim8\,{\rm d}$. Future studies should therefore concentrate on this parameter range. 

Irregular RV variations of unknown origin with amplitudes of up to $\sim180\,{\rm km\,s^{-1}}$ on timescales of years, days and even hours have been detected in some He-sdO stars. They might be connected to irregular photometric variations in some cases.

\begin{acknowledgements}

T.R.M. acknowledges support from the UK's Science and Technology Facilities Council, grant ST/L000733/1. E.Z. and C.H. are supported by the Deutsche Forschungsgemeinschaft (DFG) through grants HE1356/45-2 and HE1356/62-1. T.K. acknowledges support by the Netherlands Research School for Astronomy (NOVA). V.S. is supported by Deutsches Zentrum f\"ur Luft- und Raumfahrt (DLR) under grant 50\,OR\,1110. 

Based on observations at the La Silla-Paranal Observatory of the European Southern Observatory for programmes number 165.H-0588(A), 079.D-0288(A), 080.D-0685(A), 081.D-0819, 082.D-0649, 084.D-0348, 089.D-0265(A), 090.D-0012(A), 091.D-0038(A) and 092.D-0040(A). 

Based on observations collected at the Centro Astron\'omico Hispano Alem\'an (CAHA) at Calar Alto, operated jointly by the Max-Planck Institut f\"ur Astronomie and the Instituto de Astrof\'isica de Andaluc\'ia (CSIC). 

Based on observations with the William Herschel and Isaac Newton Telescopes operated by the Isaac Newton Group at the Observatorio del Roque de los Muchachos of the Instituto de Astrofisica de Canarias on the island of La Palma, Spain.

Based on observations with the Southern Astrophysical Research (SOAR) telescope operated by the U.S. National Optical Astronomy Observatory (NOAO), the Ministério da Ciencia e Tecnologia of the Federal Republic of Brazil (MCT), the University of North Carolina at Chapel Hill (UNC), and Michigan State University (MSU). 

Based on observations obtained at the Gemini Observatory, which is operated by the Association of Universities for Research in Astronomy, Inc., under a cooperative agreement
with the NSF on behalf of the Gemini partnership: the National Science Foundation (United States), the Science and Technology Facilities Council (United Kingdom), the
National Research Council (Canada), CONICYT (Chile), the Australian Research Council (Australia), Ministério da Ciência e Tecnologia (Brazil) 
and Ministerio de Ciencia, Tecnología e Innovación Productiva  (Argentina). 

Funding for the SDSS and SDSS-II has been provided by the Alfred P. Sloan Foundation, the Participating Institutions, the National Science Foundation, the U.S. Department of Energy, the National Aeronautics and Space Administration, the Japanese Monbukagakusho, the Max Planck Society, and the Higher Education Funding Council for England. The SDSS Web Site is http://www.sdss.org/.

The SDSS is managed by the Astrophysical Research Consortium for the Participating Institutions. The Participating Institutions are the American Museum of Natural History, Astrophysical Institute Potsdam, University of Basel, University of Cambridge, Case Western Reserve University, University of Chicago, Drexel University, Fermilab, the Institute for Advanced Study, the Japan Participation Group, Johns Hopkins University, the Joint Institute for Nuclear Astrophysics, the Kavli Institute for Particle Astrophysics and Cosmology, the Korean Scientist Group, the Chinese Academy of Sciences (LAMOST), Los Alamos National Laboratory, the Max-Planck-Institute for Astronomy (MPIA), the Max-Planck-Institute for Astrophysics (MPA), New Mexico State University, Ohio State University, University of Pittsburgh, University of Portsmouth, Princeton University, the United States Naval Observatory, and the University of Washington. 

\end{acknowledgements}

\begin{table*}
\caption{\label{tab1} Parameters of 139 hydrogen-rich hot subdwarfs (89 RV variable, 50 RV variable candidates). {Solved binaries are marked in bold face and their orbital parameters can be found in Kupfer et al. (\cite{kupfer15}) and references therein.}}
\begin{tabular}{lllllllllll}
\hline\hline
\noalign{\smallskip}
Name & Class & $m_{V}$ & $T_{\rm eff}$ & $\log{g}$ & $\log{y}$ & $d$   & $\Delta t$ & $\Delta RV_{\rm max}$        & N & $\log{p}$ \\
     &       & [mag]   & [K]           &           &           & [kpc] & [d]        & ${\rm [km\,s^{-1}]}$ &    &    \\
\noalign{\smallskip}
\hline
\noalign{\smallskip}
{\bf J082332.09+113641.9}\tablefootmark{b} & sdB   &  16.7 & $31200 \pm  600$  &  $5.79 \pm 0.06$ & $-2.0 \pm 0.1$  & $ 2.6   _{-0.2 }  ^{+0.2}$  &   53.9447  &  $359.0  \pm   6.5$  & 22 & $<-680$ \\    
{\bf J113840.68-003531.7}\tablefootmark{c} & sdB   &  14.5 & $31200 \pm  600$  &  $5.54 \pm 0.09$ & $<-3.0       $  & $ 1.2   _{-0.1 }  ^{+0.2}$  & 3361.5592  &  $332.0 \pm   14.0$  & 31 & $<-680$ \\
{\bf J165404.26+303701.8}\tablefootmark{c} & sdB   &  15.4 & $24900 \pm  800$  &  $5.39 \pm 0.12$ & $-2.4 \pm 0.1$  & $ 1.8   _{-0.3 }  ^{+0.3}$  &    2.9365  &  $271.0 \pm   17.0$  & 38 & $<-680$ \\     
{\bf J225638.34+065651.1}\tablefootmark{c} & sdB   &  15.3 & $28500 \pm  500$  &  $5.64 \pm 0.05$ & $-2.3 \pm 0.2$  & $ 1.5   _{-0.1 }  ^{+0.1}$  &   42.3494  &  $269.0 \pm   14.0$  & 50 & $<-680$ \\     
{\bf J172624.10+274419.3}\tablefootmark{c} & sdB   &  16.0 & $32600 \pm  500$  &  $5.84 \pm 0.05$ & $-2.2 \pm 0.1$  & $ 1.9   _{-0.1 }  ^{+0.1}$  &   55.9741  &  $263.0 \pm   12.0$  & 38 & $<-680$ \\
{\bf J150513.52+110836.6}\tablefootmark{c} & sdB   &  15.4 & $33200 \pm  500$  &  $5.80 \pm 0.10$ & $-2.3 \pm 0.1$  & $ 1.5   _{-0.2 }  ^{+0.2}$  &   43.6564  &  $222.0 \pm    8.0$  & 42 & $<-680$ \\
{\bf J134632.66+281722.7}\tablefootmark{b} & sdB   &  14.9 & $28800 \pm  600$  &  $5.46 \pm 0.07$ & $-2.6 \pm 0.2$  & $ 1.6   _{-0.1 }  ^{+0.2}$  &    0.9988  &  $191.0 \pm    7.0$  & 41 & $<-680$ \\
{\bf J002323.99-002953.2}\tablefootmark{c} & sdB   &  15.5 & $29200 \pm  500$  &  $5.69 \pm 0.05$ & $-2.0 \pm 0.1$  & $ 1.6   _{-0.1 }  ^{+0.1}$  &    1.0413  &  $168.0 \pm    4.0$  & 47 & $<-680$ \\     
{\bf J083006.17+475150.4}\tablefootmark{b} & sdB   &  16.0 & $25300 \pm  600$  &  $5.38 \pm 0.06$ & $<-3.0       $  & $ 2.5   _{-0.2 }  ^{+0.2}$  & 4405.6747  &  $164.0 \pm    9.0$  & 37 & $<-680$ \\     
{\bf J095238.93+625818.9}\tablefootmark{b} & sdB   &  14.8 & $27700 \pm  600$  &  $5.59 \pm 0.06$ & $-2.6 \pm 0.1$  & $ 1.2   _{-0.1 }  ^{+0.1}$  & 1183.7390  &  $154.0 \pm    8.0$  & 34 & $<-680$ \\     
{\bf J162256.66+473051.1}\tablefootmark{d} & sdB   &  16.2 & $29000 \pm  600$ &   $5.65 \pm 0.06$ & $-1.9 \pm 0.1$  & $ 2.3   _{-0.2 }  ^{+0.2}$  &    1.9832  &  $135.0 \pm    4.5$  & 34 & $<-680$ \\     
{\bf J012022.94+395059.4}\tablefootmark{e} & sdB   &  15.4 & $28500 \pm  100$  &  $5.42 \pm 0.01$ & $-3.0 \pm 0.1$  & $ 2.1   _{-0.0 }  ^{+0.0}$  & 1358.9782  &  $129.0 \pm    6.5$  & 22 & $<-680$ \\     
J173606.25+315842.7 & sdB   &  17.0 & $31300 \pm  300$  &  $5.87 \pm 0.09$ & $-2.5 \pm 0.2$  & $ 2.8   _{-0.3 }  ^{+0.3}$  & 1567.7104  &  $195.0 \pm   12.0$  & 12 & $-537.49$ \\         
{\bf J032138.67+053840.0}\tablefootmark{b} & sdB   &  15.0 & $30700 \pm  500$  &  $5.74 \pm 0.06$ & $-2.4 \pm 0.1$  & $ 1.3   _{-0.1 }  ^{+0.1}$  & 1699.1435  &  $110.0 \pm    9.0$  & 46 & $-536.46$ \\     
J191908.76+371423.9 & sdB   &  17.2 & $28300 \pm  400$  &  $5.69 \pm 0.10$ & $-2.7 \pm 0.3$  & $ 3.4   _{-0.4 }  ^{+0.5}$  &   68.8608  &  $237.0 \pm   12.0$  & 15 & $-526.21$ \\     
{\bf J102151.64+301011.9}\tablefootmark{a} & sdB   &  18.3 & $30700 \pm  500$  &  $5.71 \pm 0.06$ & $<-3.0       $  & $ 5.8   _{-0.5 }  ^{+0.5}$  &   14.9363  &  $277.0 \pm   37.0$  & 19 & $-508.16$ \\        
{\bf J204613.40-045418.7}\tablefootmark{c} & sdB   &  16.2 & $31600 \pm  500$  &  $5.54 \pm 0.08$ & $<-3.0       $  & $ 2.8   _{-0.3 }  ^{+0.3}$  &  286.2265  &  $259.0 \pm   16.0$  & 22 & $-480.28$ \\     
J173806.51+451701.7 & sdB   &  17.4 & $30500 \pm  500$  &  $5.40 \pm 0.08$ & $<-3.0       $  & $ 5.5   _{-0.6 }  ^{+0.6}$  &    1.9536  &  $233.0 \pm    8.5$  & 13 & $-461.58$ \\     
{\bf J183249.04+630910.7}\tablefootmark{b} & sdB   &  15.8 & $26800 \pm  700$  &  $5.29 \pm 0.09$ & $-2.6 \pm 0.1$  & $ 2.7   _{-0.3 }  ^{+0.4}$  & 1487.7733  &  $141.0 \pm    8.0$  & 17 & $-453.62$  \\     
J164326.04+330113.1\tablefootmark{a} & sdB   &  16.3 & $27900 \pm  500$  &  $5.62 \pm 0.07$ & $-2.3 \pm 0.2$  & $ 2.4   _{-0.2 }  ^{+0.2}$  &    2.8085  &  $175.0 \pm    5.5$  & 10 & $-452.26$ \\     
{\bf J011857.19-002546.5}\tablefootmark{b} & sdB   &  14.9 & $27900 \pm  600$  &  $5.55 \pm 0.07$ & $<-3.0       $  & $ 1.3   _{-0.1 }  ^{+0.1}$  &  265.2187  &  $140.0 \pm    8.0$  & 43 & $-386.44$ \\     
J192059.78+372220.0\tablefootmark{f} & sdB   &  15.8 & $27600 \pm  600$  &  $5.40 \pm 0.10$ & $-2.5 \pm 0.3$  & $ 2.4   _{-0.3 }  ^{+0.3}$  &    1.9589  &  $123.0 \pm    4.5$  & 15 & $-319.72$ \\     
{\bf J150829.02+494050.9}\tablefootmark{b} & sdB   &  17.7 & $29600 \pm  600$  &  $5.73 \pm 0.07$ & $-2.3 \pm 0.1$  & $ 4.3   _{-0.4 }  ^{+0.5}$  & 2161.9292  &  $209.0 \pm   15.5$  & 58 & $-269.10$ \\     
J180940.41+234328.4 & sdB   &  16.5 & $28500 \pm  300$  &  $5.44 \pm 0.06$ & $-2.9 \pm 0.2$  & $ 3.3   _{-0.3 }  ^{+0.3}$  & 2909.9029  &  $342.0 \pm   29.5$  & 36 & $-215.11$ \\     
J183349.79+652056.3 & sdB   &  17.4 & $27200 \pm  500$  &  $5.56 \pm 0.12$ & $-2.6 \pm 0.1$  & $ 4.1   _{-0.6 }  ^{+0.7}$  &   68.8591  &  $177.0 \pm    9.5$  & 16 & $-190.20$ \\     
{\bf J095101.28+034757.0}\tablefootmark{b} & sdB   &  15.9 & $29800 \pm  300$  &  $5.48 \pm 0.04$ & $-2.8 \pm 0.3$  & $ 2.5   _{-0.1 }  ^{+0.1}$  &    1.0425  &  $183.0 \pm   13.0$  & 31 & $-170.79$ \\     
{\bf J082053.53+000843.4}\tablefootmark{g} & sdB   &  15.2 & $26700 \pm  900$  &  $5.48 \pm 0.10$ & $-2.0 \pm 0.1$  & $ 1.6   _{-0.2 }  ^{+0.3}$  &  388.9794  &  $ 99.0 \pm   11.5$  & 24 & $-153.08$ \\     
J080738.96-083322.6 & sdB   &  17.2 & $27600 \pm  600$  &  $5.61 \pm 0.17$ & $-2.7 \pm 0.3$  & $ 3.6   _{-0.7 }  ^{+0.9}$  &    0.0736  &  $298.0 \pm   19.0$  & 27 & $-136.25$ \\    
{\bf J152222.15-013018.3}\tablefootmark{b} & sdB   &  17.8 & $25200 \pm  700$  &  $5.47 \pm 0.09$ & $<-3.0       $  & $ 5.2   _{-0.6 }  ^{+0.7}$  &    3.0055  &  $173.0 \pm   30.0$  & 26 & $-126.07$ \\     
J155628.34+011335.0\tablefootmark{a} & sdB   &  16.2 & $32700 \pm  600$  &  $5.51 \pm 0.08$ & $-2.9 \pm 0.2$  & $ 3.1   _{-0.3 }  ^{+0.4}$  & 4412.8910  &  $118.0 \pm   10.5$  & 15 & $-121.64$ \\     
{\bf J113241.58-063652.8}\tablefootmark{b} & sdO   &  16.2 & $46400 \pm 1000$ &   $5.89 \pm 0.07$ & $-2.9 \pm 0.2$  & $ 2.4   _{-0.2 }  ^{+0.2}$  & 1517.8240  &  $103.0 \pm   10.0$  & 32 & $-108.89$ \\  
J222850.00+391917.4 & sdB   &  16.4 & $33500 \pm  900$ &   $5.80 \pm 0.10$ & $-1.7 \pm 0.1$  & $ 2.4   _{-0.3 }  ^{+0.4}$  & 2051.8410  &  $104.0 \pm    7.5$  & 40 & $-85.63$ \\     
J173057.94+320737.0 & sdB   &  16.2 & $28200 \pm  700$  &  $5.40 \pm 0.05$ & $-2.9 \pm 0.2$  & $ 3.0   _{-0.2 }  ^{+0.2}$  &    1.9680  &  $ 94.0 \pm    8.0$  &  6 & $-69.42$ \\     
J083334.76-045759.4 & sdB   &  18.2 & $20500 \pm  700$  &  $5.52 \pm 0.10$ & $<-3.0       $  & $ 5.0   _{-0.7 }  ^{+0.8}$  &   14.8908  &  $161.0 \pm    8.5$  & 11 & $-66.11$ \\     
J164853.26+121703.0 & sdB   &  18.5 & $30400 \pm  500$ &   $5.38 \pm 0.11$ & $<-3.0       $  & $ 9.3   _{-1.2 }  ^{+1.4}$  &    0.0684  &  $135.0 \pm   13.0$  & 11 & $-64.89$ \\     
J072245.27+305233.4 & sdB   &  18.0 & $25900 \pm  700$  &  $5.61 \pm 0.16$ & $-2.6 \pm 0.2$  & $ 5.0   _{-0.9 }  ^{+1.2}$  &    1.0019  &  $123.0 \pm   12.0$  &  7 & $-62.09$ \\     
J093059.63+025032.3 & sdB   &  15.0 & $30000 \pm  600$  &  $5.67 \pm 0.18$ & $-2.7 \pm 0.2$  & $ 1.3   _{-0.3 }  ^{+0.3}$  & 2986.7695  &  $ 91.0 \pm    9.0$  & 10 & $-49.22$ \\     
J203526.46+141948.4 & sdB   &  18.7 & $30200 \pm  600$  &  $5.57 \pm 0.07$ & $-2.9 \pm 0.2$  & $ 8.3   _{-0.8 }  ^{+0.9}$  &    1.0235  &  $163.0 \pm   25.5$  & 12 & $-33.12$ \\
J203843.97+141706.0 & sdOB  &  18.7 & $36800 \pm 1000$  &  $5.89 \pm 0.20$ & $-2.4 \pm 0.3$  & $ 6.8   _{-1.5 }  ^{+1.9}$  &    0.9067  &  $102.0 \pm   10.5$  & 12 & $-32.22$ \\     
J095229.62+301553.6\tablefootmark{a} & sdOB  &  18.5 & $35200 \pm 1200$  &  $5.05 \pm 0.17$ & $<-3.0       $  & $16.0   _{-3.3 }  ^{+3.8}$  & 1155.7612  &  $198.0 \pm   27.5$  &  5 & $-28.52$ \\     
J154531.01+563944.7 & sdB   &  17.0 & $26200 \pm  900$  &  $5.13 \pm 0.14$ & $-2.0 \pm 0.2$  & $ 5.5   _{-1.0 }  ^{+1.2}$  & 2527.7769  &  $ 70.0 \pm    8.5$  &  9 & $-27.76$ \\     
J200959.27-115519.9 & sdB   &  18.7 & $29700 \pm  700$  &  $5.31 \pm 0.08$ & $<-3.0       $  & $10.7   _{-1.2 }  ^{+1.3}$  &    1.9832  &  $156.0 \pm   23.0$  &  8 & $-27.48$ \\
J005107.01+004232.5 & sdOB  &  15.9 & $38500 \pm  300$ &  $5.83  \pm 0.07$ & $-1.0 \pm 0.1$  & $ 2.0   _{-0.2 }  ^{+0.2}$  &    2.0256  &  $ 54.0 \pm    6.5$  &  7 & $-24.96$ \\
J104248.94+033355.3 & sdO   &  17.6 & $41200 \pm 3200$ &  $4.83  \pm 0.15$ & $-2.1 \pm 0.4$  & $14.5   _{-2.8 }  ^{+3.4}$  & 2246.6948  &  $ 49.0 \pm    5.0$  &  2 & $-24.34$ \\     
J181141.86+241902.7 & sdB   &  18.7 &      $-$          &        $-$       &      $-$        &              $-$            &    0.9972  &  $248.0 \pm   26.0$  &  7 & $-23.56$ \\     
J071424.12+401645.9 & sdB   &  18.2 & $27700 \pm  700$  &  $5.38 \pm 0.11$ & $-2.6 \pm 0.1$  & $ 7.6   _{-1.1 }  ^{+1.2}$  &    2.9312  &  $152.0 \pm   24.0$  &  9 & $-23.37$ \\ 
J204300.90+002145.0\tablefootmark{a} & sdO   &  17.9 & $40200 \pm  700$  &  $6.15 \pm 0.13$ & $-1.3 \pm 0.4$  & $ 3.6   _{-0.5 }  ^{+0.6}$  &   18.8480  &  $ 65.0 \pm    6.5$  &  9 & $-22.54$ \\        
J191645.87+371224.5 & sdB   &  18.3 & $33200 \pm 1000$ &   $5.84 \pm 0.17$ & $-2.7 \pm 0.2$  & $ 5.6   _{-1.1 }  ^{+1.4}$  &    3.0338  &  $134.0 \pm   23.5$  & 19 & $-22.15$ \\     
J094750.71+162731.8 & sdB   &  17.4 & $30000 \pm  700$ &   $6.25 \pm 0.31$ & $-2.2 \pm 0.3$  & $ 2.1   _{-0.7 }  ^{+1.0}$  &    0.8902  &  $130.0 \pm   13.5$  &  5 & $-20.08$ \\     
J115358.81+353929.0\tablefootmark{a} & sdOB  &  16.6 & $29400 \pm  500$  &  $5.49 \pm 0.06$ & $-2.5 \pm 0.3$  & $ 3.3   _{-0.3 }  ^{+0.3}$  & 1151.6544  &  $ 79.0 \pm    9.5$  &  5 & $-19.15$ \\     
J175125.67+255003.5\tablefootmark{a} & sdB   &  17.4 & $30600 \pm  500$  &  $5.48 \pm 0.08$ & $<-3.8       $  & $ 5.0   _{-0.5 }  ^{+0.6}$  & 1533.6229  &  $ 72.0 \pm   10.0$  &  8 & $-16.50$ \\  
J125702.30+435245.8\tablefootmark{a} & sdB   &  18.2 & $28000 \pm 1100$ &  $5.77  \pm 0.17$ & $<-3.0       $  & $ 4.9   _{-1.0 }  ^{+1.3}$  &    0.0098  &  $ 63.0 \pm   16.5$  &  3 & $-16.32$ \\  
J165446.26+182224.6 & sdB   &  18.6 & $30100 \pm  500$ &  $5.50  \pm 0.08$ & $-1.7 \pm 0.1$  & $ 8.5   _{-0.9 }  ^{+1.0}$  & 1396.0335  &  $ 48.0 \pm    5.5$  &  3 & $-15.27$ \\     
J120855.51+403716.1 & sdB   &  18.6 & $34100 \pm  900$  &  $5.98 \pm 0.13$ & $-1.5 \pm 0.1$  & $ 5.4   _{-0.9 }  ^{+1.0}$  &    0.0260  &  $171.0 \pm   20.0$  &  7 & $-14.61$ \\     
J164122.32+334452.0 & sdB   &  15.5 & $28200 \pm  500$  &  $5.49 \pm 0.11$ & $-2.5 \pm 0.3$  & $ 1.9   _{-0.3 }  ^{+0.3}$  & 2213.5393  &  $ 77.0 \pm    8.0$  &  8 & $-14.60$ \\     
J211421.39+100411.4 & sdOB  &  18.4 & $36100 \pm  900$  &  $5.48 \pm 0.13$ & $-2.5 \pm 0.3$  & $ 9.2   _{-1.4 }  ^{+1.6}$  & 1427.1132  &  $ 69.0 \pm   12.0$  &  7 & $-14.02$ \\     
J170810.97+244341.6\tablefootmark{a} & sdOB  &  18.5 & $35600 \pm  800$  &  $5.58 \pm 0.14$ & $-0.8 \pm 0.1$  & $ 8.5   _{-1.4 }  ^{+1.6}$  &    0.0125  &  $160.0 \pm   26.0$  &  3 & $-13.73$ \\     
J153411.10+543345.2\tablefootmark{a} & sdOB  &  16.9 & $34800 \pm  700$  &  $5.64 \pm 0.09$ & $-2.6 \pm 0.3$  & $ 3.8   _{-0.4 }  ^{+0.5}$  &    0.0184  &  $ 83.0 \pm   18.5$  &  8 & $-12.52$ \\     
J224518.65+220746.5 & sdB   &  16.6 & $34000 \pm  800$  &  $5.82 \pm 0.07$ & $-2.2 \pm 0.1$  & $ 2.6   _{-0.3 }  ^{+0.3}$  & 1080.8857  &  $ 70.0 \pm   11.5$  &  9 & $-12.28$ \\     
J120613.40+205523.1 & sdOB  &  18.4 & $35000 \pm  500$  &  $5.35 \pm 0.07$ & $<-3.0       $  & $10.5   _{-0.9 }  ^{+1.0}$  &    2.9112  &  $ 91.0 \pm   23.5$  & 10 & $-11.37$ \\     
J204247.51+001913.9\tablefootmark{h} & sdB   &  19.6 & $34200 \pm  400$  &  $5.89 \pm 0.08$ & $-1.3 \pm 0.1$  & $ 9.6   _{-1.0 }  ^{+1.1}$  & 1393.1941  &  $ 69.0 \pm   10.0$  &  3 & $-10.83$ \\   
J151314.23+234248.8 & sdB   &  17.1 & $28700 \pm  300$ &  $5.69  \pm 0.10$ & $-2.3 \pm 0.2$  & $ 3.3   _{-0.4 }  ^{+0.4}$  &    2.0006  &  $ 58.0 \pm    8.5$  &  3 & $-10.83$ \\     
J082944.75+132302.5 & sdOB  &  17.2 & $39700 \pm  600$  &  $5.42 \pm 0.04$ & $<-3.0       $  & $ 6.1   _{-0.3 }  ^{+0.3}$  &   24.9992  &  $ 90.0 \pm   16.5$  &  5 & $-10.40$ \\       
\noalign{\smallskip}
\hline\hline
\end{tabular}
\end{table*}

\begin{table*}
\begin{tabular}{lllllllllll}
\hline\hline
\noalign{\smallskip}
Name & Class & $m_{V}$ & $T_{\rm eff}$ & $\log{g}$ & $\log{y}$ & $d$   & $\Delta t$ & $\Delta RV_{\rm max}$        & N & $\log{p}$ \\
     &       & [mag]   & [K]           &           &           & [kpc] & [d]        & ${\rm [km\,s^{-1}]}$ &    &    \\
\noalign{\smallskip}
\hline
\noalign{\smallskip}
J074534.16+372718.5\tablefootmark{a} & sdB   &  17.9 & $37500 \pm  500$  &  $5.90 \pm 0.09$ & $<-3.0       $  & $ 4.6   _{-0.5 }  ^{+0.5}$  &    0.0363  &  $ 64.0 \pm   17.0$  &  8 & $-9.74$ \\     
J202313.83+131254.9\tablefootmark{a} & sdB   &  17.2 & $29600 \pm  600$  &  $5.64 \pm 0.14$ & $-2.1 \pm 0.1$  & $ 3.8   _{-0.6 }  ^{+0.7}$  & 1201.7981  &  $123.0 \pm   19.0$  &  5 & $-9.20$ \\     
J162610.34+130401.6 & sdB   &  19.4 & $33900 \pm  500$ &  $5.63  \pm 0.10$ & $-1.0 \pm 0.1$  & $12.1   _{-1.5 }  ^{+1.7}$  &  780.7541  &  $ 51.0 \pm    8.0$  &  3 & $-9.16$ \\
J030607.95+382335.7\tablefootmark{i} & sdO   &  16.8 & $30100 \pm  300$ &  $5.64  \pm 0.03$ & $-2.1 \pm 0.1$  & $ 3.2   _{-0.1 }  ^{+0.1}$  & 2210.7452  &  $ 48.0 \pm    6.5$  &  8 & $-8.85$ \\         
J204451.08-062753.8 & sdO   &  20.0 & $57100 \pm 5200$ &  $5.61  \pm 0.15$ & $-2.2 \pm 0.4$  & $21.4   _{-4.2 }  ^{+5.1}$  & 1087.0571  &  $ 62.0 \pm   10.5$  &  3 & $-7.88$ \\     
J091615.49+132833.1 & sdB   &  17.5 & $30900 \pm  400$ &  $5.48  \pm 0.05$ & $<-3.0       $  & $ 5.4   _{-0.4 }  ^{+0.4}$  &    0.9512  &  $ 55.0 \pm   11.5$  &  3 & $-7.58$ \\ 
J163413.09+163109.5 & sdB   &  18.3 & $34600 \pm  900$ &  $4.73  \pm 0.12$ & $-2.0 \pm 0.5$  & $20.7   _{-3.1 }  ^{+3.5}$  & 1105.3751  &  $ 21.0 \pm    5.5$  &  3 & $-7.44$ \\       
J123220.09+260913.3 & sdB   &  18.1 & $33700 \pm 1100$ &   $5.40 \pm 0.16$ & $-1.3 \pm 0.2$  & $ 8.5   _{-1.7 }  ^{+2.0}$  &    1.0302  &  $134.0 \pm   27.0$  &  5 & $-7.36$ \\     
J185129.02+182358.8 & sdB   &  16.8 & $27800 \pm  700$ &   $5.38 \pm 0.10$ & $<-3.0       $  & $ 3.9   _{-0.5 }  ^{+0.6}$  &    0.0808  &  $105.0 \pm   18.0$  & 22 & $-7.33$ \\
J220048.67+123612.4\tablefootmark{h} & sdO   &  18.6 & $64200 \pm 2600$ &  $5.63  \pm 0.11$ & $-1.3 \pm 0.1$  & $11.4   _{-1.6 }  ^{+1.8}$  & 2437.2535  &  $ 53.0 \pm    9.5$  &  3 & $-7.04$ \\     
J153752.95+160201.8 & sdB   &  18.4 & $32300 \pm  500$  &  $5.47 \pm 0.07$ & $<-3.0       $  & $ 8.5   _{-0.8 }  ^{+0.9}$  &    0.0361  &  $ 68.0 \pm   12.5$  &  3 & $-7.03$ \\         
J183229.22+402418.4 & sdO   &  15.7 & $40100 \pm  600$ &  $5.35  \pm 0.11$ & $-2.0 \pm 0.2$  & $ 3.3   _{-0.4 }  ^{+0.5}$  &    3.0098  &  $ 50.0 \pm   11.5$  &  5 & $-6.82$ \\     
J181126.83+233413.7 & sdB   &  18.4 &      $-$          &        $-$       &      $-$        &              $-$            &  1.0156  &  $121.0 \pm   20.5$  &  7 & $-6.47$ \\     
J204448.63+153638.8\tablefootmark{a}  & sdB   &  17.9 & $29600 \pm  600$  &  $5.57 \pm 0.09$ & $-2.2 \pm 0.1$  & $ 5.7   _{-0.7 }  ^{+0.7}$  &    3.0489  &  $101.0 \pm   17.5$  &  7 & $-6.41$ \\     
J185414.11+175200.2 & sdOB  &  16.9 & $35200 \pm  700$  &  $5.89 \pm 0.08$ & $-1.4 \pm 0.1$  & $ 2.9   _{-0.3 }  ^{+0.3}$  &    6.0874  &  $ 81.0 \pm   22.0$  & 10 & $-6.25$ \\     
J171629.92+575121.2\tablefootmark{a} & sdOB  &  18.2 & $37500 \pm  800$  &  $5.57 \pm 0.10$ & $<-0.7       $  & $ 7.8   _{-0.9 }  ^{+1.0}$  & 3195.9096  &  $ 67.0 \pm   15.5$  & 12 & $-6.14$ \\     
J184434.74+412158.7 & sdB   &  17.3 & $27200 \pm  500$ &  $5.57  \pm 0.12$ & $-2.6 \pm 0.1$  & $ 4.0   _{-0.6 }  ^{+0.7}$  &    2.9795  &  $ 56.0 \pm   14.0$  &  5 & $-5.72$ \\     
J091136.73+124015.2 & sdB   &  18.2 &      $-$          &        $-$       &      $-$        &              $-$            &    0.0173  &  $ 75.0 \pm   16.5$  &  3 & $-5.31$ \\     
J151337.80+195012.5 & sdB   &  18.9 &      $-$          &        $-$       &      $-$        &              $-$            &    0.0354  &  $ 98.0 \pm   33.5$  &  4 & $-5.16$ \\     
J172727.55+091215.5\tablefootmark{i} & sdO   &  17.5 & $40100 \pm 1100$ &  $5.36  \pm 0.09$ & $<-2.1       $  & $ 7.4   _{-0.8 }  ^{+0.9}$  &    0.0141  &  $ 55.0 \pm   10.5$  &  6 & $-5.10$ \\ 
J112242.69+613758.5\tablefootmark{a} & sdB   &  15.4 & $29300 \pm  500$  &  $5.69 \pm 0.10$ & $-2.3 \pm 0.3$  & $ 1.5   _{-0.2 }  ^{+0.2}$  &    0.0469  &  $ 83.0 \pm   18.5$  &  6 & $-5.08$ \\     
J161140.50+201857.0\tablefootmark{a} & sdOB  &  18.5 & $36900 \pm  700$  &  $5.89 \pm 0.13$ & $-1.2 \pm 0.1$  & $ 6.1   _{-0.9 }  ^{+1.1}$  &    0.9472  &  $108.0 \pm   23.5$  &  5 & $-4.77$ \\     
J065044.30+383133.7 & sdOB  &  17.3 & $34200 \pm  400$  &  $5.76 \pm 0.07$ & $-2.9 \pm 0.2$  & $ 3.9   _{-0.3 }  ^{+0.4}$  &    0.0131  &  $ 88.0 \pm   13.5$  & 14 & $-4.63$ \\     
J170645.57+243208.6\tablefootmark{a} & sdB   &  17.8 & $32000 \pm  500$ &  $5.59  \pm 0.07$ & $<-4.0       $  & $ 5.5   _{-0.5 }  ^{+0.6}$  &    0.0125  &  $ 46.0 \pm   12.0$  &  3 & $-4.41$ \\     
J083359.65-043521.9 & sdOB  &  18.3 & $36100 \pm  500$  &  $5.92 \pm 0.11$ & $-1.9 \pm 0.2$  & $ 5.5   _{-0.7 }  ^{+0.8}$  &   14.9765  &  $ 88.0 \pm   25.5$  & 11 & $-4.39$ \\     
J140545.25+014419.0\tablefootmark{a} & sdB   &  15.8 & $27300 \pm  800$ &  $5.37  \pm 0.16$ & $-1.9 \pm 0.2$  & $ 2.5   _{-0.5 }  ^{+0.6}$  &    0.0263  &  $ 25.0 \pm    8.0$  &  3  & $-4.12$ \\     
\noalign{\smallskip}
\hline
\noalign{\smallskip}
J160534.96+062733.5 & sdB   &  19.3 &      $-$          &        $-$       &      $-$        &              $-$            &   1.0113  &  $132.0 \pm   41.0$  &  8 & {\it $-$3.97} \\         
J221920.67+394603.5 & sdO   &  17.3 & $47000 \pm 3500$  &  $5.73 \pm 0.16$ & $<-3.0       $  & $ 4.7   _{-0.9 }  ^{+1.2}$  &   62.8679  &  $ 66.0 \pm   12.5$  &  8 & {\it $-$3.93} \\     
J183840.52+400226.8 & sdB   &  17.8 & $29300 \pm  900$  &  $5.52 \pm 0.13$ & $-1.6 \pm 0.2$  & $ 5.5   _{-0.9 }  ^{+1.1}$  &    2.9795  &  $ 74.0 \pm   20.0$  &  5 & {\it $-$3.89} \\     
J115716.37+612410.7\tablefootmark{a} & sdB   &  17.2 & $29900 \pm  500$  &  $5.59 \pm 0.08$ & $-3.2 \pm 0.8$  & $ 4.0   _{-0.4 }  ^{+0.5}$  & 2250.6902  &  $102.0 \pm   27.0$  &  7 & {\it $-$3.63} \\     
J113303.70+290223.0\tablefootmark{a} & sdB/DA&  18.9 &      $-$          &        $-$       &      $-$        &              $-$            &    0.0158  &  $ 95.0 \pm   30.0$  &  3  & {\it $-$3.39} \\     
J161817.65+120159.6\tablefootmark{a} & sdB   &  18.0 & $32100 \pm 1000$ &   $5.35 \pm 0.23$ & $<0.0        $  & $ 8.1   _{-2.1 }  ^{+2.8}$  &    0.0427  &  $105.0 \pm   28.0$  &  4 & {\it $-$3.35} \\     
J205101.72+011259.7 & sdB+X &  17.6 &      $-$          &        $-$       &      $-$        &              $-$            &    0.0141  &  $ 91.0 \pm   31.5$  &  8 & {\it $-$3.28} \\     
J133638.81+111949.4\tablefootmark{a} & sdB   &  17.3 & $27500 \pm  500$ &  $5.49  \pm 0.08$ & $-2.7 \pm 0.2$  & $ 4.4   _{-0.5 }  ^{+0.5}$  &    0.0301  &  $ 48.0 \pm   14.0$  &  3 & {\it $-$3.25} \\ 
J094044.07+004759.6\tablefootmark{h} & sdB   &  19.1 & $37000 \pm  800$ &  $5.82  \pm 0.13$ & $-0.1 \pm 0.1$  & $ 8.8   _{-1.3 }  ^{+1.5}$  & 2982.7971  &  $ 30.0 \pm    8.5$  &  2 & {\it $-$3.24} \\         
J210454.89+110645.5\tablefootmark{a} & sdOB  &  17.3 & $37800 \pm  700$  &  $5.63 \pm 0.10$ & $-2.4 \pm 0.2$  & $ 4.9   _{-0.6 }  ^{+0.6}$  & 2548.0064  &  $139.0 \pm   27.5$  &  9 & {\it $-$3.14} \\     
J211651.96+003328.5\tablefootmark{a} & sdB   &  18.0 & $27900 \pm  800$ &  $5.78  \pm 0.15$ & $-3.9 \pm 0.7$  & $ 4.3   _{-0.8 }  ^{+0.9}$  &    0.0161  &  $ 47.0 \pm   15.0$  &  3 & {\it $-$3.08} \\     
J091428.87+125023.8 & sdB   &  18.0 & $33600 \pm  600$ &  $5.54  \pm 0.11$ & $<-3.0       $  & $ 7.0   _{-0.9 }  ^{+1.1}$  &    0.0176  &  $ 49.0 \pm   13.5$  &  3 & {\it $-$3.07} \\    
J112014.86+412127.3 & sdB   &  18.1 &      $-$          &        $-$       &      $-$        &              $-$            & 1503.8023  &  $ 23.0 \pm    7.5$  &  2 & {\it $-$2.98} \\    
J173614.19+335249.5 & sdB   &  18.8 &      $-$          &        $-$       &      $-$        &              $-$            &    0.0410  &  $ 85.0 \pm   26.0$  &  5 & {\it $-$2.97} \\     
J092520.70+470330.6\tablefootmark{a} & sdB   &  17.7 & $28100 \pm  900$ &  $5.17  \pm 0.15$ & $-2.5 \pm 0.2$  & $ 7.5   _{-1.4 }  ^{+1.7}$  &    0.0126  &  $ 40.0 \pm   12.5$  &  3 & {\it $-$2.88} \\     
J171617.33+553446.7\tablefootmark{a} & sdB   &  17.2 & $32900 \pm  900$  &  $5.48 \pm 0.09$ & $<-3.0       $  & $ 4.9   _{-0.6 }  ^{+0.7}$  &    0.0125  &  $130.0 \pm   40.5$  &  9 & {\it $-$2.85} \\     
J064809.54+380850.1 & sdB   &  18.4 & $29300 \pm  800$ &  $5.26  \pm 0.13$ & $-2.8 \pm 0.3$  & $ 9.8   _{-1.6 }  ^{+1.9}$  &    0.9989  &  $ 48.0 \pm   13.0$  &  5 & {\it $-$2.85} \\
J075937.15+541022.2\tablefootmark{a} & sdB   &  17.8 & $31300 \pm  700$ &  $5.30  \pm 0.10$ & $-3.3 \pm 0.3$  & $ 7.6   _{-1.0 }  ^{+1.1}$  &    0.0233  &  $ 40.0 \pm   18.5$  &  3 & {\it $-$2.75} \\         
J001844.33-093855.0 & sdB   &  18.8 &      $-$          &        $-$       &      $-$        &              $-$            & 1169.8455  &  $ 27.0 \pm    8.0$  &  3 & {\it $-$2.75} \\     
J130439.57+312904.8\tablefootmark{a} & sdOB  &  17.1 & $38100 \pm  600$ &  $5.69  \pm 0.12$ & $-0.4 \pm 0.1$  & $ 4.1   _{-0.6 }  ^{+0.6}$  &    0.0163  &  $ 49.0 \pm   27.5$  &  3 & {\it $-$2.63} \\     
J143347.59+075416.9 & sdOB  &  16.7 & $36600 \pm  600$ &  $6.16  \pm 0.13$ & $<-0.5       $  & $ 1.9   _{-0.3 }  ^{+0.3}$  &  805.7659  &  $ 52.0 \pm   10.5$  & 11 & {\it $-$2.61} \\     
J153540.30+173458.8 & sdB   &  18.0 &      $-$          &        $-$       &      $-$        &              $-$            &    0.0168  &  $ 58.0 \pm   16.5$  &  3 & {\it $-$2.57} \\     
J202758.63+773924.5\tablefootmark{a} & sdO   &  17.9 & $46200 \pm 3200$ &   $5.48 \pm 0.18$ & $-2.8 \pm 0.9$  & $ 8.2   _{-1.8 }  ^{+2.2}$  &    1.9601  &  $114.0 \pm   33.0$  &  3 & {\it  $-$2.48} \\     
J215648.71+003620.7\tablefootmark{a}  & sdB   &  18.0 & $30800 \pm  800$  &  $5.77 \pm 0.12$ & $-2.2 \pm 0.3$  & $ 4.7   _{-0.7 }  ^{+0.8}$  &  822.1114  &  $100.0 \pm   28.0$  &  6 & {\it $-$2.38} \\     
J073701.45+225637.6 & sdB   &  16.8 & $28100 \pm  300$ &  $5.45  \pm 0.04$ & $<-3.0       $  & $ 3.7   _{-0.2 }  ^{+0.2}$  &    2.0639  &  $ 53.0 \pm   14.5$  &  5 & {\it $-$2.36} \\     
J220810.05+115913.9 & sdB   &  17.4 & $27200 \pm  600$ &  $5.23  \pm 0.07$ & $-2.3 \pm 0.3$  & $ 6.1   _{-0.6 }  ^{+0.6}$  & 2172.7020  &  $ 42.0 \pm   12.5$  &  5 & {\it $-$2.31} \\     
J172919.04+072204.5 & sdO   &  17.3 & $49200 \pm 1900$ &  $5.78  \pm 0.12$ & $-3.0 \pm 0.4$  & $ 4.6   _{-0.7 }  ^{+0.8}$  &    0.0179  &  $ 58.0 \pm   20.0$  &  5 & {\it $-$2.22} \\     
J031226.01+001018.2 & sdB   &  19.2 &      $-$          &        $-$       &      $-$        &              $-$            & 2552.8670  &  $ 71.0 \pm   30.5$  &  2 & {\it $-$2.17} \\ 
J204546.81-054355.6\tablefootmark{a} & sdB   &  17.9 & $35500 \pm  500$ &  $5.47  \pm 0.09$ & $-1.4 \pm 0.2$  & $ 7.3   _{-0.8 }  ^{+0.9}$  &    0.0128  &  $ 41.0 \pm   16.5$  &  4 & {\it $-$2.15} \\     
J133200.95+673325.7 & sdOB  &  17.2 & $37400 \pm  500$ &  $5.90  \pm 0.09$ & $-1.5 \pm 0.1$  & $ 3.4   _{-0.4 }  ^{+0.4}$  & 2584.9083  &  $ 53.0 \pm   14.5$  &  7 & {\it $-$2.09} \\     
J120427.94+172745.3 & sdB   &  18.3 & $25100 \pm  900$  &  $5.25 \pm 0.15$ & $-2.6 \pm 0.4$  & $ 8.2   _{-1.5 }  ^{+1.9}$  &    0.0282  &  $ 68.0 \pm   29.0$  &  3 & {\it $-$2.05} \\        
J204550.97+153536.3 & sdB   &  18.2 & $30300 \pm  500$ &  $5.62  \pm 0.09$ & $<-3.0       $  & $ 6.3   _{-0.7 }  ^{+0.8}$  &    5.9148  &  $ 58.0 \pm   13.5$  &  7 & {\it $-$1.98} \\     
J135807.96+261215.5\tablefootmark{a} & sdB   &  17.9 & $33500 \pm  600$  &  $5.66 \pm 0.10$ & $>+2.0       $  & $ 5.8   _{-0.7 }  ^{+0.8}$  &    0.0302  &  $ 86.0 \pm   26.0$  &  6 & {\it $-$1.89} \\ 
J113935.45+614953.9\tablefootmark{a} & sdB   &  16.9 & $28800 \pm  900$ &  $5.27  \pm 0.15$ & $-2.8 \pm 0.3$  & $ 4.9   _{-0.9 }  ^{+1.1}$  &    0.0112  &  $ 30.0 \pm   10.5$  &  3 & {\it $-$1.86} \\     
J155343.39+131330.4 & sdOB  &  18.5 & $36300 \pm  500$  &  $5.63 \pm 0.16$ & $-0.8 \pm 0.1$  & $ 8.1   _{-1.4 }  ^{+1.7}$  &    0.0160  &  $ 64.0 \pm   24.0$  &  3 & {\it $-$1.77} \\  
\noalign{\smallskip}
\hline\hline
\end{tabular}
\end{table*}

\begin{table*}
\begin{tabular}{lllllllllll}
\hline\hline
\noalign{\smallskip}
Name & Class & $m_{V}$ & $T_{\rm eff}$ & $\log{g}$ & $\log{y}$ & $d$   & $\Delta t$ & $\Delta RV_{\rm max}$        & N & $\log{p}$ \\
     &       & [mag]   & [K]           &           &           & [kpc] & [d]        & ${\rm [km\,s^{-1}]}$ &    &    \\
\noalign{\smallskip}
\hline
\noalign{\smallskip}
J082657.29+122818.1 & sdOB  &  17.1 & $36500 \pm  400$  &  $5.83 \pm 0.12$ & $-1.4 \pm 0.1$  & $ 3.4   _{-0.5 }  ^{+0.5}$  &    0.0142  &  $ 67.0 \pm   22.0$  &  4 & {\it $-$1.73} \\     
J152705.03+110843.9\tablefootmark{a} & sdOB  &  17.3 & $37600 \pm  500$ &  $5.62  \pm 0.10$ & $-0.5 \pm 0.1$  & $ 4.8   _{-0.5 }  ^{+0.6}$  &    0.0543  &  $ 43.0 \pm   12.0$  &  5 & {\it $-$1.73} \\     
J052544.93+630726.0\tablefootmark{a} & sdOB  &  17.7 & $35600 \pm  800$ &  $5.85  \pm 0.10$ & $-1.6 \pm 0.2$  & $ 4.3   _{-0.5 }  ^{+0.6}$  &    0.0264  &  $ 42.0 \pm   15.0$  &  5 & {\it $-$1.73} \\     
J100535.76+223952.1\tablefootmark{a} & sdB   &  18.4 & $29000 \pm  700$ &  $5.43  \pm 0.13$ & $-2.7 \pm 0.2$  & $ 7.9   _{-1.3 }  ^{+1.5}$  &    0.0192  &  $ 41.0 \pm   14.0$  &  4 & {\it $-$1.71} \\     
J164204.37+440303.2 & sdB   &  16.8 & $29300 \pm  800$ &  $5.09  \pm 0.13$ & $-2.5 \pm 0.3$  & $ 5.7   _{-0.9 }  ^{+1.1}$  &    0.0273  &  $ 31.0 \pm   11.5$  &  4 & {\it $-$1.68} \\     
J090957.82+622927.0 & sdO   &  16.4 & $48000 \pm 4900$ &  $5.68  \pm 0.17$ & $-1.7 \pm 0.6$  & $ 3.4   _{-0.8 }  ^{+1.0}$  &    0.0461  &  $ 37.0 \pm   12.0$  &  4 & {\it $-$1.64} \\     
J152458.81+181940.5 & sdO   &  18.3 & $52300 \pm 2500$ &  $5.28  \pm 0.08$ & $-2.8 \pm 0.3$  & $13.5   _{-1.5 }  ^{+1.7}$  &    0.0155  &  $ 41.0 \pm   15.0$  &  3 & {\it $-$1.60} \\     
J112140.20+183613.7 & sdB   &  18.6 & $28100 \pm  500$  &  $5.46 \pm 0.10$ & $-1.8 \pm 0.1$  & $ 8.3   _{-1.0 }  ^{+1.2}$  &    0.9796  &  $ 71.0 \pm   26.0$  &  4 & {\it $-$1.57} \\     
J151254.55+150447.0 & sdOB  &  17.8 & $38300 \pm  600$  &  $6.01 \pm 0.10$ & $-1.5 \pm 0.2$  & $ 4.0   _{-0.5 }  ^{+0.5}$  &    0.0229  &  $ 65.0 \pm   28.0$  &  3 & {\it $-$1.54} \\     
J233406.11+462249.3\tablefootmark{a} & sdB   &  17.7 & $34600 \pm  500$ &  $5.71  \pm 0.09$ & $-1.3 \pm 0.1$  & $ 4.9   _{-0.6 }  ^{+0.6}$  &    0.0248  &  $ 31.0 \pm   12.0$  &  3 & {\it $-$1.53} \\     
J095054.97+460405.2 & sdB   &  18.0 & $28500 \pm  500$ &  $5.24  \pm 0.07$ & $-2.3 \pm 0.3$  & $ 8.1   _{-0.8 }  ^{+0.8}$  &    0.0390  &  $ 42.0 \pm   16.5$  &  3 & {\it $-$1.52} \\     
J112526.95+112902.6 & sdOB  &  17.4 & $36100 \pm  700$  &  $6.06 \pm 0.12$ & $-0.8 \pm 0.1$  & $ 2.9   _{-0.4 }  ^{+0.5}$  &    0.0142  &  $ 70.0 \pm   31.0$  &  4 & {\it $-$1.50} \\     
J163834.68+265110.2 & sdOB  &  17.0 & $36000 \pm  300$ &  $5.80  \pm 0.05$ & $-1.6 \pm 0.1$  & $ 3.4   _{-0.2 }  ^{+0.2}$  &    0.0159  &  $ 40.0 \pm   13.0$  &  4 & {\it $-$1.50} \\     
J203017.81+131849.2 & sdOB  &  16.8 & $37100 \pm  500$ &  $5.92  \pm 0.09$ & $-1.4 \pm 0.1$  & $ 2.7   _{-0.3 }  ^{+0.3}$  & 1200.7860  &  $ 52.0 \pm   20.0$  &  5 & {\it $-$1.47} \\     
J130059.20+005711.7\tablefootmark{a} & sdOB  &  16.5 & $40700 \pm  500$ &  $5.53  \pm 0.10$ & $-0.6 \pm 0.1$  & $ 3.9   _{-0.4 }  ^{+0.5}$  &    0.0123  &  $ 36.0 \pm   14.5$  &  3 & {\it $-$1.43} \\     
\noalign{\smallskip}
\hline\hline
\end{tabular}
\tablefoot{
\tablefoottext{a}{Atmospheric parameters taken from Geier et al. (\cite{geier11a}).}
\tablefoottext{b}{Atmospheric parameters taken from Kupfer et al. (\cite{kupfer15}).}
\tablefoottext{c}{Atmospheric parameters taken from Geier et al. (\cite{geier11b}).}
\tablefoottext{d}{Atmospheric parameters taken from Schaffenroth et al. (\cite{schaffenroth14}).}
\tablefoottext{e}{Atmospheric parameters taken from \O stensen et al. (\cite{oestensen13}).}
\tablefoottext{f}{Atmospheric parameters taken from Schaffenroth et al. in prep.}
\tablefoottext{g}{Atmospheric parameters taken from Geier et al. (\cite{geier11c}).}
\tablefoottext{h}{Atmospheric parameters derived from a spectrum taken with ESO-VLT/FORS1.}
\tablefoottext{i}{Atmospheric parameters derived from a spectrum taken with WHT/ISIS.}
}
\end{table*}

\begin{table*}
\caption{\label{tab2} Parameters of 25 helium-rich hot subdwarfs (14 RV variable, 11 RV variable candidates).}
\begin{tabular}{lllllllllll}
\hline\hline
\noalign{\smallskip}
Name & Class & $m_{V}$ & $T_{\rm eff}$ & $\log{g}$ & $\log{y}$ & $d$   & $\Delta t$ & $\Delta RV_{\rm max}$        & N & $\log{p}$ \\
     &       & [mag]   & [K]           &           &           & [kpc] & [d]        & ${\rm [km\,s^{-1}]}$ &    &    \\
\noalign{\smallskip}
\hline
\noalign{\smallskip}
J232757.46+483755.2\tablefootmark{a} & He-sdO&  15.8 & $64700 \pm 2000$ &   $5.40 \pm 0.08$ & $>+2.0       $  & $ 4.2   _{-0.4 }  ^{+0.5}$  & 1799.6136  &  $176.0 \pm   20.5$  & 59 & $-680.31$ \\     
J141549.05+111213.9\tablefootmark{a} & He-sdO&  16.1 & $43100 \pm  800$  &  $5.81 \pm 0.17$ & $>+2.0       $  & $ 2.4   _{-0.4 }  ^{+0.5}$  &    0.0075  &  $125.0 \pm   17.0$  & 35 & $-86.42$ \\
J103549.68+092551.9\tablefootmark{a} & He-sdO&  16.3 & $48100 \pm  600$ &  $6.02  \pm 0.13$ & $>+2.0       $  & $ 2.2   _{-0.3 }  ^{+0.4}$  & 3541.9636  &  $ 53.0 \pm    4.0$  &  6 & $-54.25$ \\     
J170045.09+391830.3 & He-sdOB & 18.2 & $36500 \pm 1600$ &   $5.87 \pm 0.16$ & $+0.1 \pm 0.1$  & $ 5.5   _{-1.0 }  ^{+1.2}$  & 2160.0414  &  $118.0 \pm   11.5$  & 10 & $-44.76$ \\         
J161014.87+045046.6 & He-sdO&  17.3 & $48400 \pm 1400$ &   $6.31 \pm 0.09$ & $>+2.0       $  & $ 2.5   _{-0.3 }  ^{+0.3}$  &    0.0124  &  $138.0 \pm   17.0$  & 14 & $-31.77$ \\        
J110215.45+024034.1\tablefootmark{a} & He-sdO&  17.5 & $56600 \pm 4200$ &  $5.36  \pm 0.22$ & $>+2.0       $  & $ 8.9   _{-2.2 }  ^{+3.0}$  &    0.0332  &  $ 62.0 \pm    8.5$  &  3 & $-10.91$ \\
J174516.32+244348.3\tablefootmark{a} & He-sdO&  17.7 & $43400 \pm 1000$ &   $5.62 \pm 0.21$ & $>+2.0       $  & $ 6.2   _{-1.4 }  ^{+1.8}$  & 1220.5806  &  $134.0 \pm   25.5$  & 13 & $-8.81$ \\  
J160304.07+165953.8\tablefootmark{b} & He-sdO&  16.9 & $45400 \pm  300$  &  $6.10 \pm 0.07$ & $>+2.0       $  & $ 2.5   _{-0.2 }  ^{+0.2}$  &    0.9087  &  $ 71.0 \pm   18.5$  &  5 & $-8.11$ \\
J094856.95+334151.0\tablefootmark{a} & He-sdO&  17.7 & $51000 \pm 1200$  &  $5.87 \pm 0.12$ & $+1.8 \pm 0.5$  & $ 5.1   _{-0.7 }  ^{+0.8}$  &    0.0123  &  $ 74.0 \pm   14.0$  &  3 & $-7.73$ \\
J152136.25+162150.3 & He-sdO &  17.1 & $47400 \pm 1000$  &  $5.81 \pm 0.08$ & $+1.6 \pm 0.4$  & $ 4.0   _{-0.4 }  ^{+0.4}$  & 2175.9687  &  $ 77.0 \pm   24.0$  &  9 & $-5.94$ \\ 
J163416.08+221141.0 & He-sdOB& 15.5 & $38300 \pm 1400$ &  $5.65  \pm 0.26$ & $>+2.0       $  & $ 2.0   _{-0.6 }  ^{+0.8}$  &  653.3309  &  $ 35.0 \pm    6.5$  &  6 & $-5.55$ \\
J153237.94+275636.9 & He-sdO&  18.5 & $37700 \pm 1300$  &  $6.09 \pm 0.22$ & $+0.0 \pm 0.2$  & $ 5.0   _{-1.2 }  ^{+1.5}$  &    1.0012  &  $ 73.0 \pm   16.5$  &  3 & $-5.52$ \\     
J233914.00+134214.3 & He-sdO&  17.6 & $48100 \pm 1600$  &  $5.65 \pm 0.25$ & $>+2.0       $  & $ 6.0   _{-1.6 }  ^{+2.1}$  & 1451.6391  &  $ 72.0 \pm   11.8$  & 12 & $-5.11$ \\     
J173034.09+272139.8\tablefootmark{c} & He-sdO&  18.9 & $39500 \pm  700$ &  $5.83  \pm 0.17$ & $+0.1 \pm 0.1$  & $ 8.1   _{-1.5 }  ^{+1.8}$  &  698.7112  &  $ 41.0 \pm   10.0$  &  2 & $-5.00$ \\     
\noalign{\smallskip}
\hline
\noalign{\smallskip}
J170214.00+194255.1\tablefootmark{b} & He-sdO&  15.8 & $44300 \pm  600$ &  $5.79  \pm 0.11$ & $>+2.0       $  & $ 2.1   _{-0.3 }  ^{+0.3}$  & 1665.2088  &  $ 38.0 \pm   10.0$  &  5 & {\it $-$3.76} \\     
J081329.81+383326.9 & He-sdO&  17.5 & $45800 \pm  800$ &  $6.11  \pm 0.11$ & $+1.8 \pm 0.4$  & $ 3.3   _{-0.4 }  ^{+0.5}$  &    0.0175  &  $ 54.0 \pm   13.0$  &  6 & {\it $-$3.35} \\     
J204940.85+165003.6\tablefootmark{a} & He-sdO&  17.9 & $43000 \pm  700$  &  $5.71 \pm 0.13$ & $>+2.0       $  & $ 6.2   _{-0.9 }  ^{+1.1}$  &    5.9325  &  $ 84.0 \pm   18.5$  &  7 & {\it $-$3.13} \\         
J160623.21+363005.4 & He-sdOB& 18.5 & $36400 \pm  700$  &  $5.34 \pm 0.17$ & $-0.5 \pm 0.1$  & $11.3   _{-2.1 }  ^{+2.6}$  & 1414.9811  &  $ 67.0 \pm   19.5$  &  2 & {\it $-$3.04} \\     
J112414.45+402637.1\tablefootmark{a} & He-sdO&  18.0 & $47100 \pm 1000$ &  $5.81  \pm 0.23$ & $>+1.7       $  & $ 5.9   _{-1.4 }  ^{+1.9}$  &    0.0215  &  $ 62.0 \pm   18.5$  &  3 & {\it $-$2.65} \\     
J161059.80+053625.2\tablefootmark{b} & He-sdO&  17.2 & $46300 \pm  700$ &  $6.22  \pm 0.10$ & $+1.0 \pm 0.6$  & $ 2.6   _{-0.3 }  ^{+0.3}$  &  751.7674  &  $ 38.0 \pm    9.5$  &  4 & {\it $-$2.64} \\     
J151415.66-012925.2\tablefootmark{a} & He-sdO&  17.0 & $48200 \pm  500$  &  $5.85 \pm 0.08$ & $+1.7 \pm 0.4$  & $ 3.6   _{-0.3 }  ^{+0.4}$  &    3.9687  &  $ 66.0 \pm   20.5$  &  5 & {\it $-$2.58} \\     
J161938.64+252122.4 & He-sdOB& 17.5 & $35000 \pm 2000$  &  $5.80 \pm 0.33$ & $-0.4 \pm 0.2$  & $ 4.3   _{-1.5 }  ^{+2.1}$  &    0.9716  &  $ 67.0 \pm   26.0$  &  3 & {\it $-$1.81} \\         
J160450.44+051909.2 & He-sdOB & 18.5 & $38100 \pm  700$  &  $5.22 \pm 0.27$ & $+1.2 \pm 0.2$  & $13.7   _{-3.8 }  ^{+5.2}$  &    0.9736  &  $145.0 \pm   61.0$  &  8 & {\it $-$1.75} \\     
J090252.99+073533.9 & He-sdO&  17.4 & $40100 \pm  500$  &  $5.91 \pm 0.19$ & $>+2.0       $  & $ 3.7   _{-0.7 }  ^{+0.9}$  & 1612.4334  &  $ 67.0 \pm   27.0$  &  5 & {\it $-$1.65} \\     
J081304.04-071306.5 & He-sdO&  18.6 & $48200 \pm  900$ &   $5.93 \pm 0.14$ & $+1.8 \pm 0.5$  & $ 7.0   _{-1.1 }  ^{+1.3}$  &    0.9897  &  $137.0 \pm   41.0$  &  7 & {\it $-$1.50} \\
\noalign{\smallskip}
\hline\hline
\end{tabular}
\tablefoot{
\tablefoottext{a}{Atmospheric parameters taken from Geier et al. (\cite{geier11a}).}
\tablefoottext{b}{Atmospheric parameters derived from a spectrum taken with ESO-VLT/FORS1.}
\tablefoottext{c}{Atmospheric parameters derived from a spectrum taken with WHT/ISIS.}
}
\end{table*}

\begin{table*}
\caption{\label{tab3} Parameters of 13 other types of hot stars (7 RV variable, 6 RV variable candidates).}
\begin{tabular}{lllllllllll}
\hline\hline
\noalign{\smallskip}
Name & Class & $m_{V}$ & $T_{\rm eff}$ & $\log{g}$ & $\log{y}$ & $d$   & $\Delta t$ & $\Delta RV_{\rm max}$        & N & $\log{p}$ \\
     &       & [mag]   & [K]           &           &           & [kpc] & [d]        & ${\rm [km\,s^{-1}]}$ &    &    \\
\noalign{\smallskip}
\hline
\noalign{\smallskip}
J131916.15-011404.9 & BHB   &  16.4 & $17400 \pm  800$ &  $4.55  \pm 0.15$ & $-1.9 \pm 0.2$  & $ 5.9   _{-1.1 }  ^{+1.4}$  & 2888.0925  &  $ 46.0 \pm    9.0$  &  8 & $-42.10$ \\     
J164121.22+363542.7 & BHB   &  17.4 & $19300 \pm 1000$  &  $4.55 \pm 0.10$ & $-1.9 \pm 0.2$  & $ 9.9   _{-1.4 }  ^{+1.7}$  & 1035.9093  &  $ 99.0 \pm    9.0$  &  8 & $-39.13$ \\ 
J075732.18+184329.3\tablefootmark{a} & O(He) &  18.6 & $80000 \pm 2000$ &   $5.00 \pm 0.30$ & $>+2.0       $ & $29.6   _{-9.0 } ^{+12.7}$  &    0.0216  &  $107.0 \pm   22.0$  &  6 & $-30.13$ \\     
J155610.40+254640.3\tablefootmark{b} & PG\,1159 &  17.9 & $100000_{-10000}^{+15000}$  &  $5.3 \pm 0.3$       & $>+2.0$          &  $ 16.9 _{-5.6}  ^{+8.9}$ &  231.1694  &  $116.0 \pm   21.0$  & 10 & $-17.98$ \\
J201302.58-105826.1 & MS-B  &  18.5 & $16400 \pm 1400$ &  $4.30  \pm 0.27$ & $-1.3 \pm 0.2$  & $51.8   _{-16.4}  ^{+23.6}$  &    2.0155  &  $ 61.0 \pm   11.5$  &  8 & $-13.42$ \\     
J093521.39+482432.4 & O(H)  &  18.5 & $87700 \pm20000$ &  $5.68  \pm 0.16$ & $-1.0 \pm 0.3$  & $12.0   _{-3.3 }  ^{+3.7}$  & 2269.7542  &  $ 38.0 \pm    7.5$  &  2 & $-6.97$ \\     
J161253.21+060538.7 & MS-B  &  15.5 & $15700 \pm 1400$ &  $4.18  \pm 0.29$ & $-1.0 \pm 0.2$  & $14.4   _{-4.8 }  ^{+7.2}$  &  811.5968  &  $ 38.0 \pm    7.0$  & 10 & $-6.84$ \\     
\noalign{\smallskip}
\hline
\noalign{\smallskip}
J020531.40+134739.8\tablefootmark{c} & BHB   &  18.4 & $17400 \pm  700$ &  $4.26  \pm 0.13$ & $-1.7 \pm 0.2$  & $20.3   _{-3.4 }  ^{+4.0}$  & 2781.1087  &  $ 28.0 \pm    7.0$  &  3 & {\it $-$3.64} \\     
J144023.58+135454.7 & BHB   &  18.3 & $18900 \pm  700$  &  $4.50 \pm 0.15$ & $-1.9 \pm 0.3$  & $16.1   _{-3.0 }  ^{+3.6}$  &    0.0528  &  $ 78.0 \pm   24.0$  &  4 & {\it $-$3.15} \\     
J171947.87+591604.2 & MS-B  &  16.9 & $15100 \pm  600$ &  $4.10  \pm 0.19$ & $-0.9 \pm 0.2$  & $29.2   _{-6.5 }  ^{+8.3}$  & 2568.7218  &  $ 32.0 \pm    6.5$  & 10 & {\it $-$3.11} \\
J100019.98-003413.3 & O(H)  &  17.8 & $93700 \pm10700$ &   $5.88 \pm 0.10$ & $-0.6 \pm 0.2$  & $ 7.3   _{-1.1 }  ^{+1.3}$  &    3.0114  &  $135.0 \pm   28.0$  & 16 & {\it$-$2.20} \\          
J110256.32+010012.3\tablefootmark{c} & BHB   &  18.5 & $17300 \pm  800$ &  $4.32  \pm 0.14$ & $-2.1 \pm 0.2$  & $19.5   _{-3.5 }  ^{+4.3}$  & 2735.5338  &  $ 24.0 \pm    9.0$  &  3 & {\it $-$1.77} \\     
J204149.38+003555.8\tablefootmark{c} & BHB   &  19.0 & $19400 \pm 2200$ &  $4.02  \pm 0.29$ & $-2.1 \pm 0.4$  & $38.3   _{-13.4}  ^{+20.3}$ &   38.0700  &  $ 26.0 \pm   10.5$  &  3 & {\it $-$1.71} \\   
\noalign{\smallskip}
\hline\hline
\end{tabular}
\tablefoot{
\tablefoottext{a}{Atmospheric parameters taken from Werner et al. (\cite{werner14}).}
\tablefoottext{b}{Atmospheric parameters taken from Reindl et al. (\cite{reindl15}).}
\tablefoottext{c}{Atmospheric parameters derived from a spectrum taken with ESO-VLT/FORS1.}
}
\end{table*}

\begin{appendix}

\section{Appendix}

\begin{table*}
\caption{\label{app:tab1} Parameters of 19 stars with non-significant RV variations.}
\begin{tabular}{lllllllllll}
\hline\hline
\noalign{\smallskip}
Name & Class & $m_{V}$ & $T_{\rm eff}$ & $\log{g}$ & $\log{y}$ & $d$   & $\Delta t$ & $\Delta RV_{\rm max}$        & N & $\log{p}$ \\
     &       & [mag]   & [K]           &           &           & [kpc] & [d]        & ${\rm [km\,s^{-1}]}$ &    &    \\
\noalign{\smallskip}
\hline
\noalign{\smallskip}
J085727.65+424215.4\tablefootmark{a} & He-sdO&  18.5 & $39500 \pm 1900$ &   $5.63 \pm 0.24$ & $+0.2 \pm 0.2$  & $ 8.7   _{-2.2 }  ^{+3.0}$  &    0.0657  &  $111.0 \pm   39.5$  &  4 & {\it $-$1.26} \\
J074551.13+170600.3 & sdOB  &  17.1 & $35600 \pm  400$  &  $5.54 \pm 0.05$ & $-2.8 \pm 0.1$  & $ 4.7   _{-0.3 }  ^{+0.3}$  &    9.9390  &  $ 65.0 \pm   12.0$  & 18 & {\it $-$1.26} \\     
J110445.01+092530.9\tablefootmark{a} & sdOB  &  16.3 & $35900 \pm  800$ &  $5.41  \pm 0.07$ & $-2.1 \pm 0.4$  & $ 3.8   _{-0.3 }  ^{+0.4}$  &    0.0396  &  $ 34.0 \pm   12.0$  &  4 & {\it $-$1.25} \\     
J012739.35+404357.8\tablefootmark{a} & sdO   &  16.8 & $48300 \pm 3200$ &  $5.67  \pm 0.10$ & $-1.3 \pm 0.2$  & $ 4.1   _{-0.6 }  ^{+0.7}$  &    0.0369  &  $ 45.0 \pm   17.0$  &  8 & {\it $-$1.23} \\     
J172816.87+074839.0 & sdB   &  18.4 & $30700 \pm  700$  &  $5.37 \pm 0.09$ & $-2.5 \pm 0.4$  & $ 9.0   _{-1.1 }  ^{+1.2}$  &    1.9962  &  $ 75.0 \pm   34.0$  &  7 & {\it $-$1.11} \\   
J143153.05-002824.3\tablefootmark{a} & sdOB  &  18.1 & $37300 \pm  800$  &  $6.02 \pm 0.16$ & $-0.8 \pm 0.1$  & $ 4.4   _{-0.8 }  ^{+0.9}$  &    0.0120  &  $ 64.0 \pm   20.5$  &  8 & {\it $-$1.05} \\     
J225150.80-082612.7\tablefootmark{b} & BHB   &  18.4 & $19000 \pm  500$ &  $4.98  \pm 0.09$ & $-1.8 \pm 0.3$  & $ 9.5   _{-1.1 }  ^{+1.3}$  & 2411.2964  &  $ 20.0 \pm    7.0$  &  5 & {\it $-$1.04} \\ 
J074806.15+342927.7 & sdOB  &  17.3 & $35100 \pm  800$ &  $5.72  \pm 0.08$ & $-1.7 \pm 0.1$  & $ 4.3   _{-0.5 }  ^{+0.5}$  &    5.9453  &  $ 42.0 \pm   12.5$  & 12 & {\it $-$0.95} \\     
J111225.70+392332.7 & sdOB  &  17.6 & $37800 \pm  500$ &   $5.76 \pm 0.11$ & $-0.6 \pm 0.1$  & $ 4.9   _{-0.6 }  ^{+0.7}$  &    0.0563  &  $104.0 \pm   28.0$  & 13 & {\it $-$0.92} \\         
J134352.14+394008.3\tablefootmark{a} & He-sdOB& 18.2 & $36000 \pm 2100$ &  $4.78  \pm 0.30$ & $-0.2 \pm 0.2$  & $18.8   _{-6.1 }  ^{+8.5}$  &    0.0224  &  $ 53.0 \pm   27.0$  &  3 & {\it $-$0.89} \\     
J163702.78-011351.7\tablefootmark{a}  & He-sdO&  17.3 & $46100 \pm  700$  &  $5.92 \pm 0.22$ & $>+2.0       $  & $ 3.8   _{-0.9 }  ^{+1.1}$  &    0.0853  &  $100.0 \pm   42.5$  & 12 & {\it $-$0.85} \\     
J174442.35+263829.9 & sdOB  &  17.9 &      $-$          &        $-$       &      $-$        &              $-$            &   0.0384  &  $ 88.0 \pm   44.0$  &  7 & {\it $-$0.84} \\     
J180757.08+230133.0 & He-sdO&  17.1 & $42700 \pm 1000$ &  $6.04  \pm 0.21$ & $>+2.0       $  & $ 2.9   _{-0.7 }  ^{+0.8}$  &    0.9992  &  $ 39.0 \pm   19.0$  &  4 & {\it $-$0.83} \\                                                            
J204623.12-065926.8 & O(H)   &  17.7 & $79500 \pm12500$ &  $5.74  \pm 0.13$ & $-1.1 \pm 0.2$  & $ 7.6   _{-1.6 }  ^{+1.9}$  & 1376.1081  &  $ 47.0 \pm   18.0$  &  5 & {\it $-$0.64} \\     
J075818.49+102742.5 & sdOB  &  16.4 & $37400 \pm  600$ &  $5.51  \pm 0.05$ & $<-3.0       $  & $ 3.6   _{-0.2 }  ^{+0.2}$  &    0.0596  &  $ 32.0 \pm   12.5$  &  6 & {\it $-$0.57} \\     
J215053.84+131650.5 & sdB+X &  17.0 &      $-$          &        $-$       &      $-$        &              $-$            &    0.0154  &  $ 24.0 \pm   13.5$  &  4 & {\it $-$0.56} \\    
J215307.34-071948.3 & sdB   &  17.1 & $33100 \pm 1300$ &  $5.74  \pm 0.15$ & $-2.0 \pm 0.2$  & $ 3.6   _{-0.7 }  ^{+0.8}$  &   24.9831  &  $ 50.0 \pm   27.5$  & 13 & {\it $-$0.42} \\     
J113418.00+015322.1\tablefootmark{a} & sdB   &  17.7 & $29700 \pm 1200$ &  $4.83  \pm 0.16$ & $<-4.0       $  & $11.8   _{-2.4 }  ^{+2.9}$  &    0.0757  &  $ 46.0 \pm   20.0$  &  6 & {\it $-$0.42} \\     
J170716.53+275410.4 & sdB   &  16.7 & $30200 \pm 1400$ &  $5.62  \pm 0.16$ & $<-3.0       $  & $ 3.1   _{-0.6 }  ^{+0.8}$  &    0.0124  &  $ 52.0 \pm   23.0$  &  9 & {\it $-$0.21} \\     
\noalign{\smallskip}
\hline\hline
\end{tabular}
\tablefoot{
\tablefoottext{a}{Atmospheric parameters taken from Geier et al. (\cite{geier11a}).}
\tablefoottext{b}{Atmospheric parameters derived from a spectrum taken with ESO-VLT/FORS1.}
}
\end{table*}

\end{appendix}

\end{document}